\documentclass[useAMS,usenatbib]{mn2e-accepted}

\usepackage{natbib}
\usepackage{graphicx}
\usepackage{txfonts}
\usepackage{url}
\usepackage{times}

\def\ms{\,m\,s$^{-1}$}         
\def\kms{\,km\,s$^{-1}$}       
\def\mstar{$M_*$}		
\def\rstar{$R_*$}		
\def\densstar{$\rho_*$}		
\def\mplanet{$M_{\rm P}$}	
\def\rplanet{$R_{\rm P}$}	
\def\densplanet{$\rho_{\rm P}$}
\def\mjup{$M_{\rm Jup}$}	
\def\rjup{$R_{\rm Jup}$}	
\def\teql{$T_{\rm eql}$}
\def\teff{$T_{\rm eff}$}
\def\feh{[Fe/H]}
\def\logg{$\log g_*$}
\def\vsini{$v \sin I$}
\def\mictrb{$\xi_{\rm t}$}

\def\halpha{$H_\alpha$}
\def\kms{km\, s$^{-1}$}

\def\ecos{$e \cos \omega$}
\def\esin{$e \sin \omega$}
\def\secos{$\sqrt{e} \cos \omega$}
\def\sesin{$\sqrt{e} \sin \omega$}
\def\chisq{$\chi^2$}
\def\deg{$^\circ$}
\def\rhk{$\log R'_{\rm HK}$}

\newcommand{\aap}{A\&A}
\newcommand{\apj}{ApJ}
\newcommand{\apjl}{ApJ}
\newcommand{\mnras}{MNRAS}

\newcommand{\nat}{Nature}
\newcommand{\pasp}{PASP}
\newcommand{\aapr}{A\&A~Rev.}
\newcommand{\aaps}{A\&AS}
\newcommand{\aj}{AJ}

\newcommand{\leftcell}[1]{\multicolumn{1}{l}{#1}}

\title[WASP-69b, WASP-70Ab and WASP-84b]{Three newly-discovered sub-Jupiter-mass planets: WASP-69b \& WASP-84b transit active K dwarfs and WASP-70Ab transits the evolved primary of a G4+K3 binary\thanks{Based on 
observations made with the WASP-South (South Africa) and SuperWASP-North (La Palma) 
photometric survey instruments, the RISE camera on the 2-m Liverpool Telescope under 
program PL12B13, and, all located at La Silla: the 60-cm TRAPPIST photometer, EulerCam 
and the CORALIE spectrograph, both mounted on the 1.2-m Euler-Swiss telescope, and the 
HARPS spectrograph on the ESO 3.6-m telescope under program 89.C-0151.}\thanks{The photometric time-series and radial-velocity data used in this work are available at the CDS via anonymous ftp to cdsarc.u-strasbg.fr (130.79.128.5) or via http://cdsarc.u-strasbg.fr/viz-bin/qcat?J/MNRAS/445/1114}}

\author[D.~R.~Anderson et al.]{D.~R.~Anderson,$^{1}$\thanks{E-mail: d.r.anderson@keele.ac.uk}
A.~Collier~Cameron,$^{2}$ 
L.~Delrez,$^{3}$ 
A.~P.~Doyle,$^{1}$ 
F.~Faedi,$^{4}$ 
A.~Fumel,$^{3}$ 
\newauthor
M.~Gillon,$^{3}$ 
Y.~G\'omez Maqueo Chew,$^{4}$ 
C.~Hellier,$^{1}$ 
E.~Jehin,$^{3}$ 
M.~Lendl,$^{5}$ 
P.~F.~L.~Maxted,$^{1}$ 
\newauthor
F.~Pepe,$^{5}$
D.~Pollacco,$^{4}$
D.~Queloz,$^{5,6}$ 
D.~S\'egransan,$^{5}$
I.~Skillen,$^{7}$ 
B.~Smalley,$^{1}$ 
\newauthor
A.~M.~S.~Smith,$^{1,8}$ 
J.~Southworth,$^1$
A.~H.~M.~J.~Triaud,$^{5,9}$
O.~D.~Turner,$^{1}$ 
S.~Udry$^{5}$
\newauthor
and R.~G.~West$^{4}$\\
$^1$Astrophysics Group, Keele University, Staffordshire ST5 5BG, UK.\\
$^2$SUPA, School of Physics and Astronomy, University of St. Andrews, 
           North Haugh, Fife KY16 9SS, UK\\
$^3$Institut d'Astrophysique et de G\'eophysique,  Universit\'e de 
           Li\`ege,  All\'ee du 6 Ao\^ut, 17,  Bat.  B5C, Li\`ege 1, Belgium\\
$^4$Department of Physics, University of Warwick, Coventry CV4 7AL, UK\\
$^5$Observatoire de Gen\`eve, Universit\'e de Gen\`eve, 51 Chemin 
           des Maillettes, 1290 Sauverny, Switzerland\\
$^6$Cavendish Laboratory, J J Thomson Avenue, Cambridge, CB3 0HE, UK\\
$^7$Isaac Newton Group of Telescopes, Apartado de Correos 321, E-38700 Santa Cruz de la Palma, Tenerife, Spain\\
$^8$N. Copernicus Astronomical Centre, Polish Academy of Sciences, Bartycka 18, 00-716, Warsaw, Poland \\
$^9$Department of Physics, and Kavli Institute for Astrophysics and Space Research, Massachusetts Institute of Technology, Cambridge, MA 02139, USA. 
}

\begin{document}

\date{Accepted 2014 August 22. Received 2014 August 18; in original form 2013 October 21}

\pagerange{\pageref{firstpage}--\pageref{lastpage}} \pubyear{2013}

\maketitle

\label{firstpage}

\begin{abstract}
We report the discovery of the transiting exoplanets WASP-69b, WASP-70Ab and WASP-84b, each of which orbits a bright star ($V\sim10)$. 
WASP-69b is a bloated Saturn-mass planet (0.26\,\mjup, 1.06\,\rjup) in a 3.868-d period around an active, $\sim$1-Gyr, mid-K dwarf.
ROSAT detected X-rays 60$\pm$27\arcsec\ from WASP-69. If the star is the source then the planet could be undergoing mass-loss at a rate of $\sim$10$^{12}$\,g\,s$^{-1}$. This is 1--2 orders of magnitude higher than the evaporation rate estimated for HD\,209458b and HD\,189733b, both of which have exhibited anomalously-large Lyman-$\alpha$ absorption during transit.
WASP-70Ab is a sub-Jupiter-mass planet (0.59\,\mjup, 1.16\,\rjup) in a 3.713-d orbit around the primary of a spatially-resolved, 9--10-Gyr, G4+K3 binary, with a separation of 3\farcs3 ($\geq$800\,AU). 
WASP-84b is a sub-Jupiter-mass  planet (0.69\,\mjup, 0.94\,\rjup) in an 8.523-d orbit around an active, $\sim$1-Gyr, early-K dwarf. Of the transiting planets discovered from the ground to date, WASP-84b has the third-longest period.
For the active stars WASP-69 and WASP-84, we pre-whitened the radial velocities using a low-order harmonic series. We found this reduced the residual scatter more than did the oft-used method of pre-whitening with a fit between residual radial velocity and bisector span. The system parameters were essentially unaffected by pre-whitening. 
\end{abstract}



\section{Introduction}

Radial-velocity (RV) surveys tend to focus on inactive stars because of the inherent difficulty in determining whether the source of an RV signal is stellar activity or reflex motion induced by an orbiting body (e.g. \citealt{2001A&A...379..279Q, 2003A&A...406..373S, 2004A&A...420L..27D, 2008A&A...489L...9H}). 
Transit surveys are not affected by this ambiguity as stellar activity does not produce transit-like features in lightcurves. The separation of the RV contributions due to reflex motion from those due to activity can prove simple or unnecessary in the case of short-period, giant transiting planets (e.g. \citealt{2011PASP..123..547M, 2012MNRAS.422.1988A}), but it is non-trivial for lower mass planets such as the super-Earth CoRoT-7b (\citealt{2009A&A...506..303Q, 2010A&A...520A..53L, 2011MNRAS.411.1953P, 2011A&A...531A.161F, 2011ApJ...743...75H}). 

Transiting planets orbiting one component of a multiple system have been known for some time (e.g. WASP-8; \citealt{2010A&A...517L...1Q}) and the {\it Kepler} mission recently found circumbinary transiting planets (e.g. Kepler-16 and Kepler-47; \citealt{2011Sci...333.1602D,2012Sci...337.1511O}). 
For those systems with a spatially-resolved secondary of similar brightness to the primary (e.g. WASP-77; \citealt{2013PASP..125...48M}), the secondary can be used as the reference source in ground-based observations of exoplanet atmospheres. Without a suitable reference such observations tend to either fail to reach the required precision or give ambiguous results (e.g. \citealt{2010Natur.463..637S, 2011ApJ...728...18M, 2011ASPC..450...55A}). 

The longitudinal coverage of ground-based transit surveys such as HAT-Net and HAT-South makes them relatively sensitive to longer-period planets \citep{2004PASP..116..266B, 2013PASP..125..154B}. 
For example, in the discovery of HAT-P-15b, the planet with the longest period (10.9\,d) of those found by ground-based transit surveys, transits were observed asynchronously from Arizona and Hawai'i \citep{2010ApJ...724..866K}. 
By combining data from multiple seasons, surveys such as SuperWASP can increase their sensitivity to longer periods without additional facility construction costs \citep{2006PASP..118.1407P}.

We report here the discovery of three exoplanet systems, each comprising of a giant planet transiting a bright host star ($V\sim10)$. 
WASP-69b is a bloated Saturn-mass planet in a 3.868-d period around an active mid-K dwarf. 
WASP-70Ab is a sub-Jupiter-mass planet in a 3.713-d orbit around the primary of a spatially-resolved 
G4+K3 binary. 
WASP-84b is a sub-Jupiter-mass planet in an 8.523-d orbit around an active early-K dwarf.
In Section~\ref{sec:obs} we report the photometric and spectroscopic observations 
leading to the identification, confirmation and characterisation of the exoplanet 
systems. 
In Section~\ref{sec:stars} we present spectral analyses of the host stars and, 
in Section~\ref{sec:rot}, searches of the stellar lightcurves for activity-rotation 
induced modulation.
In Section~\ref{sec:syspar} we detail the derivation of the systems' parameters from 
combined analyses. 
We present details of each system in Sections~\ref{w69-sys}, \ref{w70-sys} 
and \ref{w84-sys} and, finally, we summarise our findings in Section~\ref{summary}.


\section{Observations}
\label{sec:obs}

The WASP (Wide Angle Search for Planets) photometric survey 
\citep{2006PASP..118.1407P} monitors bright stars ($V$ = 8--15) 
using two eight-camera arrays, each with a field of view of 450 deg$^2$. 
Each array observes up to eight pointings per night with a cadence of 
5--10 min, and each pointing is followed for around five months per season. 
The WASP-South station \citep{2011EPJWC..1101004H} is hosted by 
the South African Astronomical Observatory and the SuperWASP-North station 
\citep{2011EPJWC..1101003F} is hosted by 
the Isaac Newton Group in the Observatorio del Roque de Los Muchachos on La Palma. 

The WASP data were processed and searched for transit signals as described in 
\citet{2006MNRAS.373..799C} and the candidate selection process was performed 
as described in \citet{2007MNRAS.380.1230C}. 
We perform a search for transits on each season of data from each camera separately and we perform combined searches on all WASP data available on each star, allowing for photometric offsets between datasets. 
This results in early detection, a useful check for spurious signals and increased sensitivity to shallow transits and long periods. 
We routinely correct WASP data for systematic effects using a combination of {\sc SysRem} and {\sc TFA} \citep{2005MNRAS.356.1466T, 2005MNRAS.356..557K}. 
{\sc TFA} is more effective at e.g. removing sinusoidal modulation, as can result from a non-axisymmetric distribution of star spots, which can otherwise swamp the transit signal. 
However, {\sc TFA} tends to suppress transits whereas {\sc SysRem} does not. 
Therefore we use {\sc SysRem}-detrended WASP lightcurves, corrected for modulation as necessary, in system characterisation (Sections~\ref{sec:rot} and \ref{sec:syspar}).

\begin{figure}
\includegraphics[width=90mm]{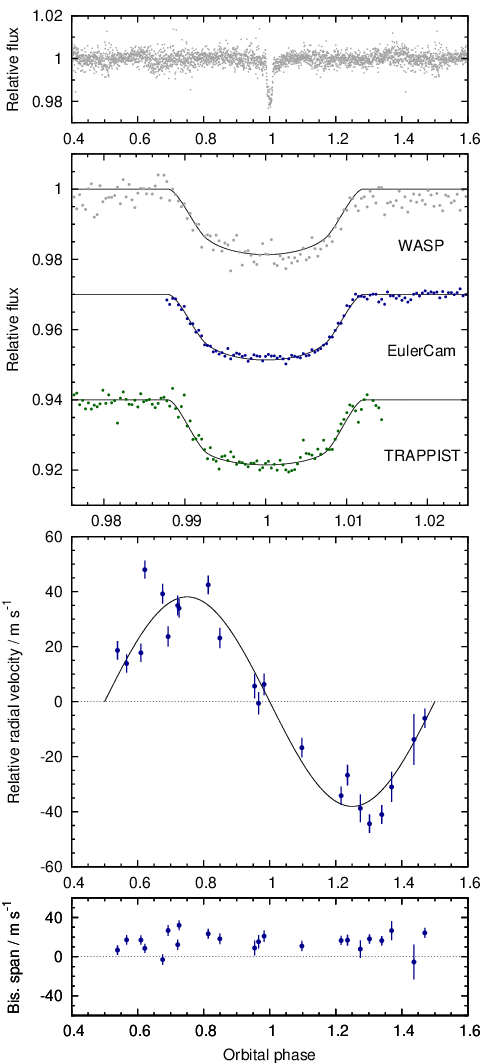}
\caption{WASP-69b discovery data. 
Top panel: WASP lightcurve folded on the transit ephemeris  
and binned in phase with a bin-width, $\Delta \phi$, equivalent to two minutes.
Second panel: Transit lightcurves from facilities as labelled, offset for clarity 
and binned with $\Delta \phi$ = 2 minutes. 
The best-fitting transit model is superimposed. 
Third panel: The pre-whitened CORALIE radial velocities with the best-fitting circular Keplerian orbit model. 
Bottom panel:  The absence of any correlation between bisector span and 
radial velocity excludes transit mimics. 
\label{fig:w69-rv-phot}} 
\end{figure}

\begin{figure}
\includegraphics[width=90mm]{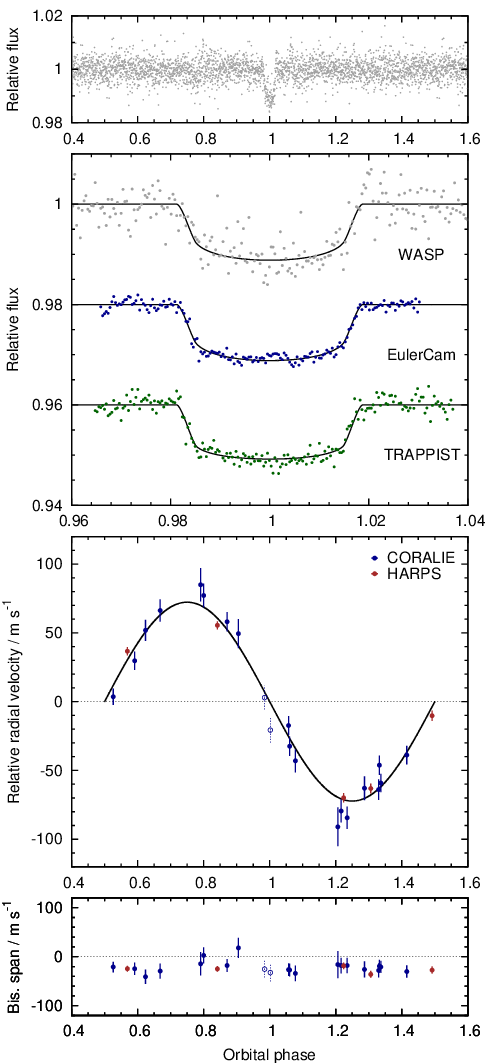}
\caption{WASP-70Ab discovery data. 
Caption as for Figure~\ref{fig:w69-rv-phot}. 
Two CORALIE RVs, taken during transit and depicted with open circles, were
excluded from the fit as we did not model the Rossiter-McLaughlin effect. 
The RVs were not pre-whitened as we did not find WASP-70A to be active. 
\label{fig:w70-rv-phot}} 
\end{figure}

\begin{figure}
\includegraphics[width=90mm]{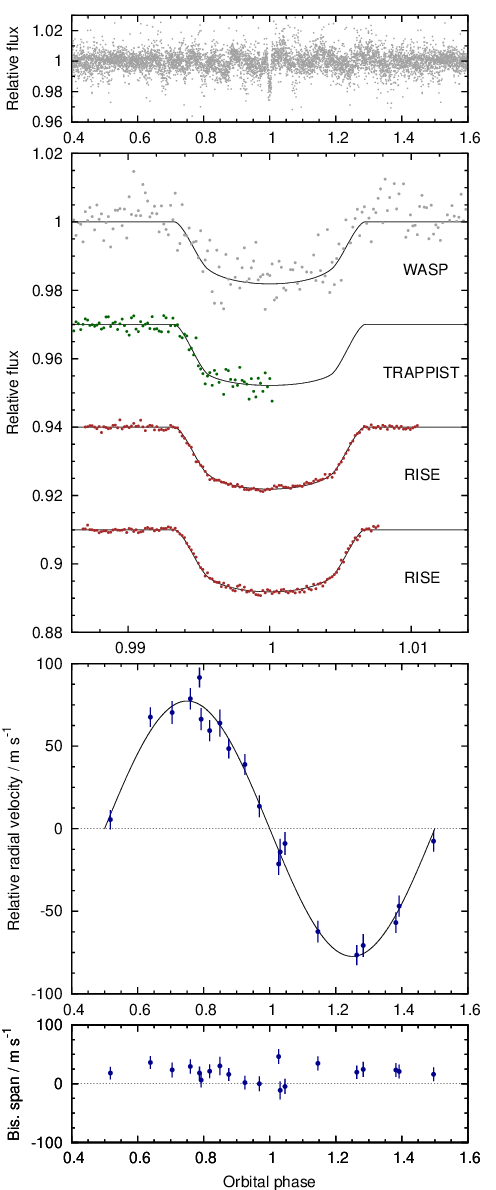}
\caption{WASP-84b discovery data. 
Caption as for Figure~\ref{fig:w69-rv-phot}. 
\label{fig:w84-rv-phot}} 
\end{figure}

We detected periodic dimmings in the WASP lightcurves of WASP-69, -70 and -84 with periods of 3.868\,d, 3.713\,d and 8.523\,d, respectively. 
We identified WASP-69b and WASP-70Ab as candidates based on data from the 2008 season alone, whereas data from two seasons (2009 and 2010) were required to pick up the longer-period WASP-84b. 
This highlights the importance of multi-epoch coverage for sensitivity to longer-period systems. 
The number of full/partial transits observed of WASP-69b was 2/2 in 2008, 2/5 in 2009 and 2/4 in 2010.
The figures for WASP-70Ab are 4/2 in 2008, 2/5 in 2009 and 2/2 in 2010. 
The figures for WASP-84b are 1/1 in 2009, 1/2 in 2010 and 2/0 in 2011. 
In the top panels of Figures~\ref{fig:w69-rv-phot}, \ref{fig:w70-rv-phot} and \ref{fig:w84-rv-phot} we plot all available WASP data, phase-folded on the best-fitting periods (Section~\ref{sec:syspar}) and corrected for systematics with {\sc SysRem}; the lightcurves of WASP-69 and WASP-84 have been corrected for rotation modulation (Section~\ref{sec:rot}). 
Coherent structure is apparent in the lightcurve of WASP-84 with a timescale of $\sim$1\,day. This is probably due to an imperfect subtraction of the rotational modulation signal and it does not appear to affect the transit (see the second panel of Figure~\ref{fig:w84-rv-phot}).
Though the structure is far reduced in a lightcurve detrended with {\sc SysRem+TFA} (not shown), we opted not to use it as the transit was also suppressed (as was evident from a combined analyis including the RISE lightcurves).

We obtained spectra of each star using the CORALIE spectrograph mounted on the 
Euler-Swiss 1.2-m telescope. 
Two spectra of WASP-84 from 2011 Dec 27 and 2012 Jan 2 were discarded as they were taken through cloud.
We also obtained five spectra of WASP-70A with the HARPS spectrograph mounted on the ESO 
3.6-m telescope. As the diameter of the HARPS fibre is half that of the CORALIE fibre 
(1\arcsec\ cf. 2\arcsec) this is a useful check for contamination of the 
CORALIE spectra by WASP-70B, located $\sim$3\arcsec\ away.  
Radial-velocity (RV) measurements were computed by weighted cross-correlation
\citep{1996A&AS..119..373B,2005Msngr.120...22P} with a numerical G2-spectral
template for WASP-70 and with a K5-spectral template for both WASP-69 and WASP-84 (Table~\ref{tab:rv}).
RV variations were detected with periods similar to those found from the WASP photometry
and with semi-amplitudes consistent with planetary-mass companions. 
The RVs are plotted, phased on the transit ephemerides, in the third panel of 
Figures~\ref{fig:w69-rv-phot}, \ref{fig:w70-rv-phot} and \ref{fig:w84-rv-phot}. 

For each star, we tested the hypothesis that the RV variations are due to
spectral-line distortions caused by a blended eclipsing binary or starspots
by performing a line-bisector analysis of the
cross-correlation functions \citep{2001A&A...379..279Q}.
The lack of correlation between bisector span and RV supports our conclusion that 
the periodic dimming and RV variation of each system are instead caused by a transiting planet (fourth panel of Figures~\ref{fig:w69-rv-phot}, \ref{fig:w70-rv-phot} and \ref{fig:w84-rv-phot}).

We performed high-quality transit observations to refine the systems' parameters using 
the 0.6-m TRAPPIST robotic telescope \citep{2011A&A...533A..88G,2011Msngr.145....2J}, EulerCam mounted on the 
1.2-m Swiss Euler telescope \citep{2012A&A...544A..72L}, and the RISE camera mounted on the 2-m Liverpool Telescope \citep{2004SPIE.5489..679S, 2008SPIE.7014E.217S}. 
Lightcurves, extracted from the images using standard aperture photometry, 
are displayed 
in the second panel of Figures~\ref{fig:w69-rv-phot}, \ref{fig:w70-rv-phot} and 
\ref{fig:w84-rv-phot} and the data are provided in Table~\ref{tab:phot}.
The RISE camera was specifically designed to obtain high-precision, high-cadence transit lightcurves and the low scatter of the WASP-84b transits demonstrate the instrument's aptness (see Figure~\ref{fig:w84-rv-phot}). 

A summary of observations is given in Table~\ref{tab:obs} and the data are 
plotted in Figures~\ref{fig:w69-rv-phot}, \ref{fig:w70-rv-phot} and 
\ref{fig:w84-rv-phot}. 
The WASP photometry, the radial-velocity measurements and the high signal-to-noise transit photometry are provided in tables online at the CDS; a guide to their content and format is  given in Tables~\ref{tab:waspphot}, \ref{tab:rv} and \ref{tab:phot}. 

\begin{table*}
\centering
\caption{Summary of observations}
\label{tab:obs}
\begin{tabular}{lcrcl}
\hline
\leftcell{Facility} & \leftcell{Date} & \leftcell{$N_{\rm obs}$} & \leftcell{$T_{\rm exp}$} & \leftcell{Filter} \\
         &      &      & \leftcell{(s)}           &        \\
\hline
{\bf WASP-69:}\\
WASP-South	    & 2008 Jun--2010 Oct	& 29\,800 & 30 & Broad (400--700 nm) \\
SuperWASP-North	    & 2008 Jun--2009 Oct	& 22\,600 & 30 & Broad (400--700 nm) \\
Euler/CORALIE	    & 2009 Aug--2011 Sep	& 22 & 1800 & Spectroscopy\\
Euler/EulerCam			& 2011 Nov 10               & 367 & $\sim$30 & Gunn-$r$ \\
TRAPPIST		    & 2012 May 21		& 873 & $\sim$10 & $z'$ \smallskip \\
{\bf WASP-70 A+B:}\\
WASP-South	    & 2008 Jun--2009 Oct	& 12\,700 & 30 & Broad (400--700 nm) \\
SuperWASP-North	    & 2008 Sep--2010 Oct	& 14\,000 & 30 & Broad (400--700 nm) \\
{\bf A:} Euler/CORALIE & 2009 Jul--2011 Oct     & 21 & 1800 & Spectroscopy\\
{\bf A:} ESO-3.6m/HARPS & 2012 Jun 26--2012 Jul 07 & 5 & 600--1800 & Spectroscopy\\
{\bf B:} Euler/CORALIE & 2009 Sep 15--22	& 4 & 1800 & Spectroscopy \smallskip \\
Euler/EulerCam			& 2011 Sep 20               & 466 & $\sim$40 & Gunn-$r$ \\
TRAPPIST			& 2011 Sep 20		& 1334 & $\sim$12 & $I+z'$ \smallskip \\
{\bf WASP-84:}\\
WASP-South	    & 2009 Jan--2011 Apr	& 14\,100 & 30 & Broad (400--700 nm) \\
SuperWASP-North	    & 2009 Dec--2011 Mar	&  8\,700 & 30 & Broad (400--700 nm) \\
Euler/CORALIE	    & 2011 Dec--2012 Mar	& 20 & 1800 & Spectroscopy \\
TRAPPIST			& 2012 Mar 01		& 557  & $\sim$25 & $z'$ \\
LT/RISE				& 2013 Jan 01       & 4322 & 4 & $V+R$ \\
LT/RISE				& 2013 Jan 18       & 1943 & 8 & $V+R$ \\
\hline
\end{tabular}
\end{table*}

\begin{table*} 
\caption{WASP photometry} 
\label{tab:waspphot} 
\begin{tabular}{lllllrrr} 
\hline 
\leftcell{Set} & \leftcell{Star} & \leftcell{Field} & \leftcell{Season} & \leftcell{Camera} & \leftcell{HJD(UTC)} & \leftcell{Mag., $M$} & \leftcell{$\sigma_{M}$}\\ 
& & & & & \leftcell{$-$2450000} & & \\
& & & & & \leftcell{(day)} & & \\
\hline
1  & WASP-69  & SW2045$-$0345 & 2008 & 223 & 4622.483160 & 9.9086  & 0.0135 \\
1  & WASP-69  & SW2045$-$0345 & 2008 & 223 & 4622.483588 & 9.8977  & 0.0139 \\
\ldots \\
10 & WASP-70 & SW2114$-$1205 & 2008 & 221 & 4622.483391	& 11.2161 & 0.0096 \\
10 & WASP-70 & SW2114$-$1205 & 2008 & 221 & 4622.483831	& 11.2098 & 0.0101 \\
\ldots \\
19 & WASP-84  & SW0846+0544   & 2011 & 226 & 5676.372824 & 10.9244 & 0.0193 \\
19 & WASP-84  & SW0846+0544   & 2011 & 226 & 5676.373264 & 10.9379 & 0.0178 \\
\hline
\end{tabular}
\begin{flushleft}
The measurements for WASP-70 include both WASP-70A and WASP-70B. 
The uncertainties are the formal errors (i.e. they have not been rescaled).
This table is available in its entirety via the CDS. 
\end{flushleft}
\end{table*}

\begin{table*}
\caption{Radial velocity measurements} 
\label{tab:rv} 
\begin{tabular}{lllrrrrr} 
\hline 
\leftcell{Set} & \leftcell{Star} & \leftcell{Spectrograph} & \leftcell{BJD(UTC)} & \leftcell{RV} & \leftcell{$\sigma_{\rm RV}$} & \leftcell{BS}\\ 
 & & & \leftcell{$-$2450000}&    &                     &   \\
 & & & \leftcell{(day)}     & \leftcell{(km s$^{-1}$)} & \leftcell{(km s$^{-1}$)} & \leftcell{(km s$^{-1}$)}\\ 
\hline
1 & WASP-69 & CORALIE & 5070.719440 & $-$9.61358 & 0.00369 & $-$0.02674\\
1 & WASP-69 & CORALIE & 5335.854399 & $-$9.64619 & 0.00376 & $-$0.01681\\
\ldots\\
2 & WASP-70A & CORALIE &  5038.745436 & $-$65.43271 & 0.00840 & $-$0.03410\\
2 & WASP-70A & CORALIE &  5098.630404 & $-$65.48071 & 0.01411 & $-$0.01614\\
\ldots\\
4 & WASP-84 & CORALIE &  6003.546822 & $-$11.52073 & 0.00822 & 0.03026\\
4 & WASP-84 & CORALIE &  6004.564579 & $-$11.56112 & 0.00661 & $-$0.00007\\
\hline
\end{tabular}
\begin{flushleft}
The presented RVs have not been pre-whitened and the uncertainties are the formal errors (i.e. with no added jitter). 
The uncertainty on bisector span (BS) is 2\,$\sigma_{\rm RV}$. 
This table is available in its entirety via the CDS. 
\end{flushleft}
\end{table*}

\begin{table*} 
\caption{EulerCam, TRAPPIST and RISE photometry} 
\label{tab:phot} 
\begin{tabular}{lllllrr} 
\hline 
\leftcell{Set} & \leftcell{Star} & \leftcell{Imager} & \leftcell{Filter} & \leftcell{BJD(UTC)} & \leftcell{Rel. flux, $F$} & \leftcell{$\sigma_{F}$}\\ 
    &      &      & \leftcell{$-$2450000} &        &                 \\
& & & \leftcell{(day)} & & \\
\hline
1	& WASP-69 & EulerCam & Gunn-r & 5845.489188 & 1.000769 & 0.000842 \\
1	& WASP-69 & EulerCam & Gunn-r & 5845.489624 & 0.999093 & 0.000840 \\
\ldots \\
4	& WASP-70A & TRAPPIST & $I+z'$ & 5825.48554  & 1.006661  & 0.002919 \\
4	& WASP-70A & TRAPPIST & $I+z'$ & 5825.48576  & 1.001043  & 0.002903 \\
\ldots \\
8	& WASP-84 & RISE & $V+R$ & 6311.743556 & 0.999610 & 0.001827 \\
8	& WASP-84 & RISE & $V+R$ & 6311.743649 & 1.000331 & 0.001829 \\
\hline
\end{tabular}
\begin{flushleft}
The flux values are differential and normalised to the out-of-transit levels. 
The contamination of the WASP-70A photometry by WASP-70B has been accounted for. 
The uncertainties are the formal errors (i.e. they have not been rescaled).
This table is available in its entirety via the CDS. 
\end{flushleft}
\end{table*}



\section{Stellar parameters from spectra}
\label{sec:stars}
The CORALIE spectra were co-added for each star to produce high-SNR spectra for analysis using the methods of \citet{2009A&A...496..259G} and \citet{2013MNRAS.428.3164D}. 
We obtained an initial estimate of effective temperature (\teff) from the \halpha\ 
line, then refined the determination using excitation balance of the Fe~{\sc i} abundance.  
The surface gravity (\logg) was determined from the Ca~{\sc i} lines at 
6162{\AA} and 6439{\AA} \citep{2010A&A...519A..51B}, along with the 
Na~{\sc i} D and Mg~{\sc i} b lines.
The elemental abundances were determined from equivalent-width measurements of 
several clean and unblended lines. 
A value for microturbulence (\mictrb) was determined from Fe~{\sc i} using the 
method of \cite{1984A&A...134..189M}. 
The quoted error estimates include that given by the uncertainties in \teff, 
\logg\ and \mictrb, as well as the scatter due to measurement and atomic data 
uncertainties.

We determined the projected stellar rotation velocity (\vsini) by fitting the 
profiles of several unblended Fe~{\sc i} lines. 
We took macroturbulence values of $2.3 \pm 0.3$\,\kms\ for WASP-70A 
and $1.2 \pm 0.3$\,\kms\ for WASP-84 from the tabulation of 
\cite{2010MNRAS.405.1907B}.
For WASP-69 and WASP-70B we assumed macroturbulence to be zero, since for mid-K 
stars it is expected to be lower than that of thermal broadening 
\citep{2008oasp.book.....G}.
An instrumental FWHM of 0.11 $\pm$ 0.01~{\AA} was determined from the telluric 
lines around 6300\AA. 

The results of the spectral analysis are given in Table~\ref{tab:stars} and 
further details of each star are given in Sections~\ref{w69-sys}, \ref{w70-sys} 
and \ref{w84-sys}.

\begin{table*}
\centering
\caption{Stellar parameters from spectra}
\begin{tabular}{ccccc} 
\hline
\leftcell{Parameter}      &   WASP-69             & WASP-70A             & WASP-70B        & WASP-84 \\ 
\hline
\teff / K      &   4700 $\pm$ 50       & 5700 $\pm$ 80        & $4900 \pm 200$  & $5300 \pm 100$    \\
\logg          &   4.5 $\pm$ 0.15      & 4.26 $\pm$ 0.09      & $4.5 \pm 0.2$   & $4.4 \pm 0.1$     \\
\mictrb\ / \kms&   0.7 $\pm$ 0.2       & 1.1 $\pm$ 0.1        & $0.7 \pm 0.2$   & $1.0 \pm 0.1$     \\
\vsini\ / \kms &   2.2 $\pm$ 0.4       & 1.8 $\pm$ 0.4        & $3.9 \pm 0.8^{\rm a}$ & $4.1 \pm 0.3$ \\
\rhk		   &   $-4.54$             & $-5.23$              &                 & $-4.43$  \smallskip \\
{[Fe/H]}       &   0.15 $\pm$ 0.08     & $-0.01 \pm 0.06$     &                 & $0.00 \pm 0.10$   \\
{[Mg/H]}       &   \ldots              & $0.05 \pm 0.04$      &                 & \ldots            \\
{[Si/H]}       &   0.42 $\pm$ 0.12     & $0.15 \pm 0.05$      &                 & $0.06 \pm 0.07$   \\
{[Ca/H]}       &   0.15 $\pm$ 0.09     & $0.07 \pm 0.10$      &                 & $0.15 \pm 0.09$   \\
{[Sc/H]}       &   0.25 $\pm$ 0.10     & $0.15 \pm 0.11$      &                 & $-0.01 \pm 0.14$  \\
{[Ti/H]}       &   0.21 $\pm$ 0.06     & $0.07 \pm 0.04$      &                 & $0.09 \pm 0.11$   \\
{[V/H]}        &   0.48 $\pm$ 0.15     & $0.02 \pm 0.08$      &                 & $0.21 \pm 0.12$   \\
{[Cr/H]}       &   0.18 $\pm$ 0.18     & $0.05 \pm 0.04$      &                 & $0.08 \pm 0.15$   \\
{[Mn/H]}       &   0.40 $\pm$ 0.07     & \ldots               &                 & \ldots\\
{[Co/H]}       &   0.42 $\pm$ 0.06     & $0.10 \pm 0.05$      &                 & $0.04 \pm 0.04$   \\
{[Ni/H]}       &   0.29 $\pm$ 0.11     & $0.05 \pm 0.05$      &                 & $-0.02 \pm 0.06$  \\
$\log A$(Li)   &   $<$ 0.05 $\pm$ 0.07 & $< 1.20 \pm 0.07$ & $< 0.71 \pm 0.25$  & $< 0.12 \pm 0.11$ \smallskip \\
Sp. Type$^{\rm b}$       &   K5                  & G4                & K3                 & K0                \\
Age$^{\rm c}$ / Gyr      & $\sim$2			     & \multicolumn{2}{c}{9--10}          & $\sim$1 \\
Distance / pc  &   50 $\pm$ 10 pc      &    \multicolumn{2}{c}{$245 \pm 20$}    & $125 \pm 20$      \\ 
Constellation  & Aquarius              & \multicolumn{2}{c}{Aquarius}           & Hydra             \\
R.A. (J2000)$^{\rm d}$	& $\rm  21^{h} 00^{m} 06\fs19$ & \multicolumn{2}{c}{$\rm 21^{h} 01^{m} 54\fs48^{\rm f}$} & $\rm  08^{h} 44^{m} 25\fs71$ \\
Dec. (J2000)$^{\rm d}$	& $\rm -05\degr 05\arcmin 40\farcs1$ & \multicolumn{2}{c}{$\rm -13\degr 25\arcmin 59\farcs8^{\rm f}$} & $\rm +01\degr 51\arcmin 36\farcs0$ \\
$B$$^{\rm d}$		& $10.93 \pm 0.06$		& \multicolumn{2}{c}{$11.75 \pm 0.13^{\rm f}$}	& $11.64 \pm 0.10$\\
$V$$^{\rm d}$		& $9.87 \pm 0.03$		& \multicolumn{2}{c}{$10.79 \pm 0.08^{\rm f}$}	& $10.83 \pm 0.08$\\
$J$$^{\rm e}$	& $8.03 \pm 0.02$	& \multicolumn{2}{c}{$10.00 \pm 0.03^{\rm f}$}	& $9.35 \pm 0.03$\\
$H$$^{\rm e}$	& $7.54 \pm 0.02$	& \multicolumn{2}{c}{$9.71 \pm 0.04^{\rm f}$}	& $8.96 \pm 0.02$\\
$K$$^{\rm e}$	& $7.46 \pm 0.02$	& \multicolumn{2}{c}{$9.58 \pm 0.03^{\rm f}$}	& $8.86 \pm 0.02$\\
2MASS J$^{\rm e}$		& 21000618$-$0505398 & \multicolumn{2}{c}{21015446$-$1325595$^{\rm f}$}	& 08442570$+$0151361\\
\hline 
\\
\end{tabular}
\label{tab:stars}
\begin{flushleft} 
$^{\rm a}$ Due to low S/N, the \vsini\ value for WASP-70B should be considered an upper limit.
\newline $^{\rm b}$ The spectral types were estimated from \teff\ using the table in \cite{2008oasp.book.....G}.
\newline $^{\rm c}$ See Sections~\ref{w69-sys}, \ref{w70-sys} and \ref{w84-sys}.
\newline $^{\rm d}$ From the Tycho-2 catalogue \citep{2000A&A...355L..27H}. 
\newline $^{\rm e}$ From the Two Micron All Sky Survey (2MASS; \citealt{2006AJ....131.1163S}).
\newline $^{\rm f}$ WASP-70A+B appear as a single entry in the Tycho-2 and 2MASS catalogues.
\end{flushleft}
\end{table*}

\section{Stellar rotation from lightcurve modulation}
\label{sec:rot}

\begin{figure*} 
\begin{center}
\includegraphics[width=0.9\textwidth]{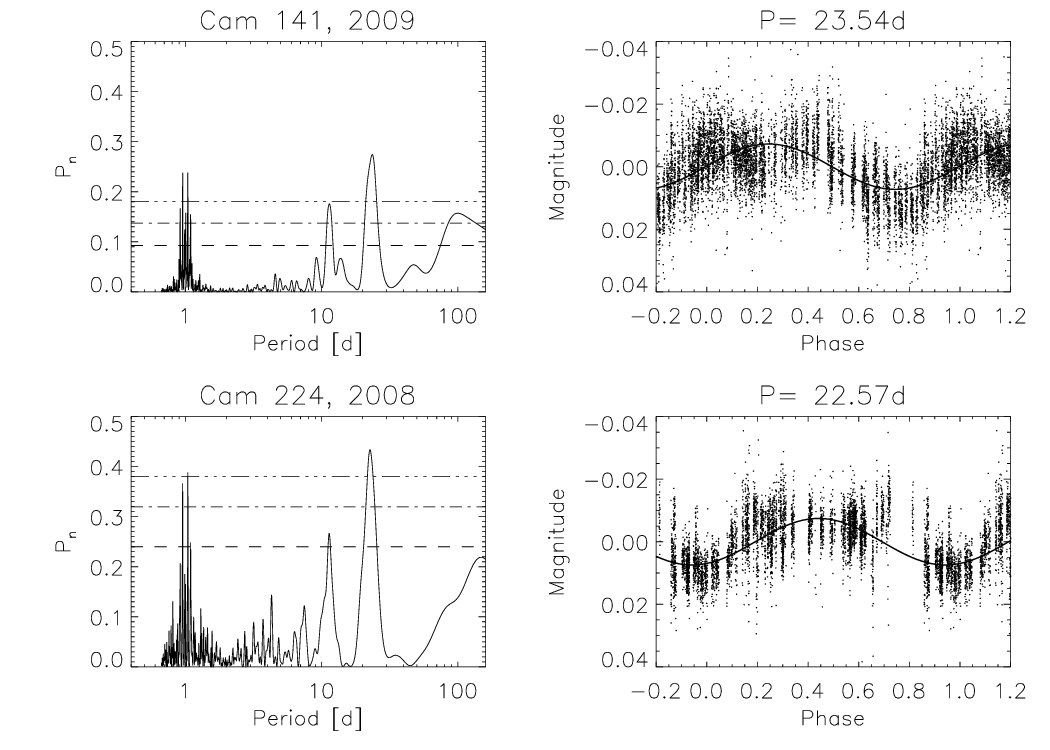}
\end{center}
\caption{{\it Left panels}: Periodograms for the WASP data from
two different observing seasons for WASP-69. Horizontal lines indicate false
alarm probability levels FAP=0.1,0.01,0.001. The WASP camera and year of
observation are noted in the titles; the first digit of the camera number denotes the observatory (1=SuperWASP-North and 2=WASP-South) and the third digit denotes the camera number.
{\it Right panels} Lightcurves folded on the best periods as noted in the
title.
\label{fig:swlomb-w69} }
\end{figure*}

\begin{figure*} 
\begin{center}
\includegraphics[width=0.9\textwidth]{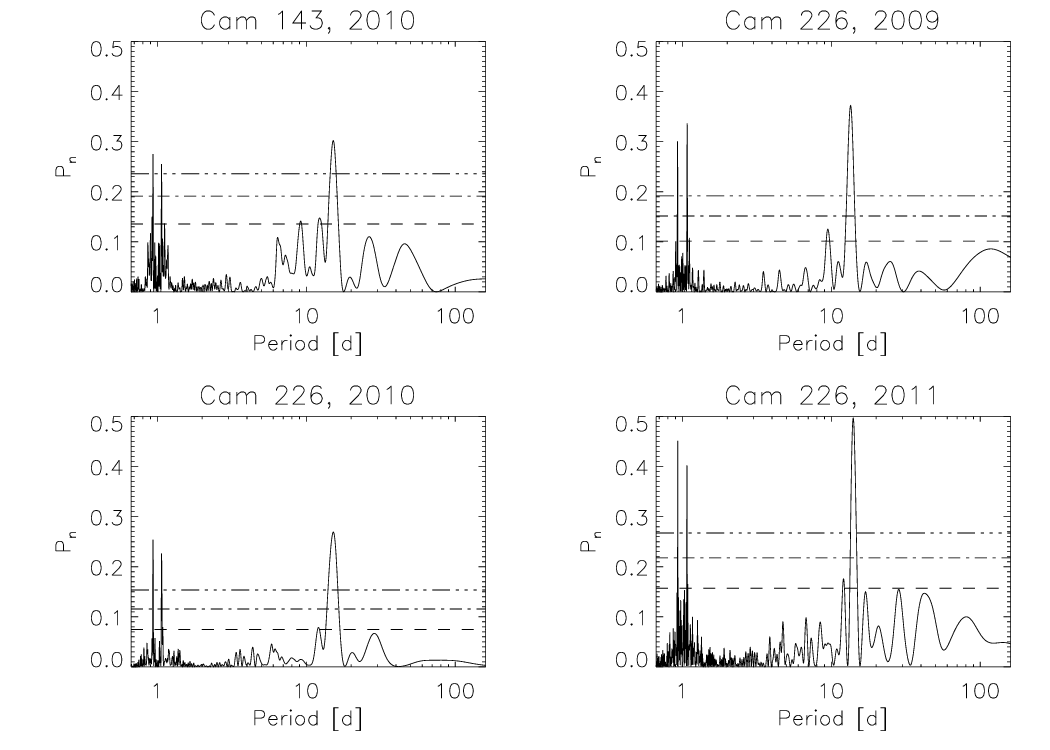}
\end{center}
\caption{Periodograms of the WASP data for WASP-84 from four independent data
sets. The WASP camera and year of observation are noted in the titles; the first digit of the camera number denotes the observatory (1=SuperWASP-North and 2=WASP-South) and the third digit denotes the camera number. 
Horizontal lines indicate false alarm probability levels
FAP=0.1,0.01,0.001.
\label{fig:swlomb-w84} }
\end{figure*}

We analysed the WASP lightcurves of each star to determine whether they show 
periodic modulation due to the combination of magnetic activity and stellar 
rotation.  
We used the sine-wave fitting method described in \citet{2011PASP..123..547M} 
to calculate periodograms such as those shown in Figures~\ref{fig:swlomb-w69} and 
\ref{fig:swlomb-w84}. 
The false alarm probability levels shown in these figures are calculated using a boot-strap Monte Carlo method also described in \citet{2011PASP..123..547M}. 

Variability due to star spots is not expected to 
be coherent on long timescales as a consequence of the finite lifetime of 
star-spots and differential rotation in the photosphere so we analysed each  
season of data separately. 
We also analysed the data from each WASP camera
separately as the data quality can vary between cameras. We
removed the transit signal from the data prior to calculating the periodograms
by subtracting a simple transit model from the lightcurve. 
We then calculated periodograms over 4\,096 uniformly spaced frequencies from 0 to 
1.5 cycles day$^{-1}$.

We found no evidence of modulation above the 1-mmag level in the WASP-70A+B lightcurves. 
The results for WASP-69 and WASP-84 are shown in Table~\ref{tab:ProtTable}. 
Taking the average of the periods for each star gives our best estimates for the 
rotation periods of $23.07\pm0.16$\,d for WASP-69 and $14.36\pm0.35$\,d for 
WASP-84.
The WASP-69 data sets obtained in 2008 with both cameras 224 and 226 
each show periods of $\sim$23/2\,days, presumably as 
a result of multiple spot groups on the surface of the star during this observing
season. The data on WASP-69 obtained by camera 223 in 2008 are affected by 
systematic instrumental noise and so are not useful for this analysis. 

\begin{figure} 
\begin{center}
\includegraphics[width=0.5\textwidth]{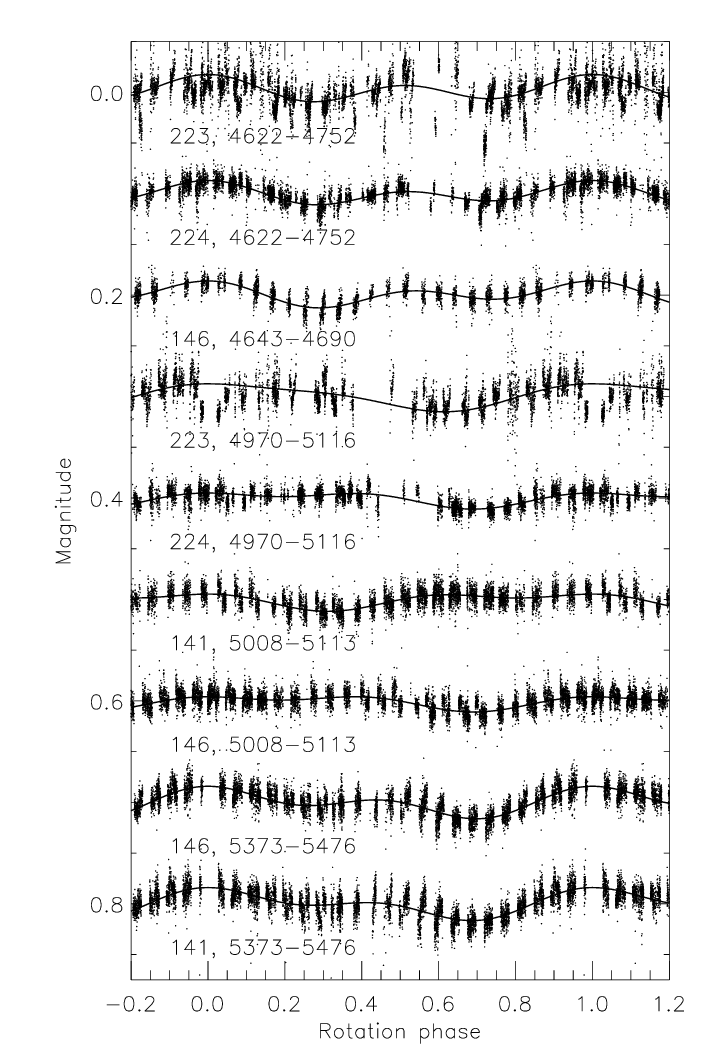}
\end{center}
\caption{WASP lightcurves of WASP-69 plotted as a function of rotation phase
for $P_{\rm rot} = 23.07$\,d. The data, detrended by {\sc SysRem}, for each 
combination of observing season and camera are offset by multiples of 0.1\,magnitudes. 
Each lightcurve is labelled with its camera number and the date range it spans, 
in truncated Julian Date: JD $-$ 2\,450\,000.
Solid lines show the
harmonic fit used to remove the rotational modulation prior to modeling the
transit. 
\label{fig:lcfit-w69} }
\end{figure}

\begin{figure} 
\begin{center}
\includegraphics[width=0.5\textwidth]{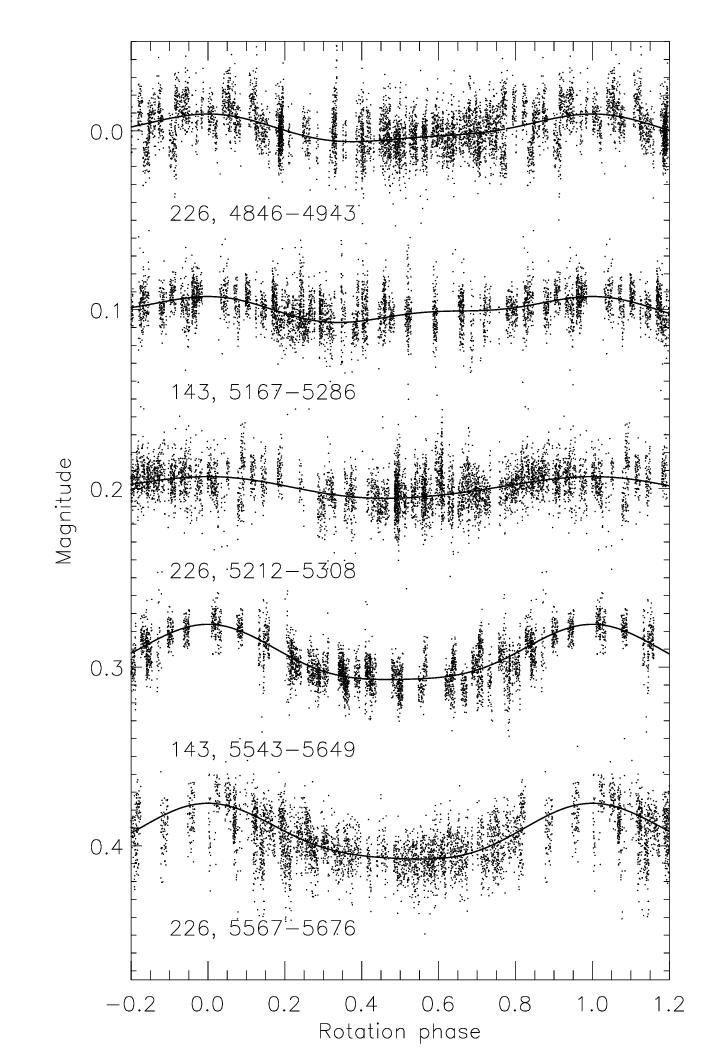}
\end{center}
\caption{WASP lightcurves of WASP-84 plotted as a function of rotation phase
for $P_{\rm rot} = 14.36$\,d. Caption as for Fig~\ref{fig:lcfit-w69}. 
\label{fig:lcfit-w84} }
\end{figure}

For both WASP-69 and WASP-84, we used a least-squares fit of the sinusoidal function and its first harmonic
to model the rotational modulation in the lightcurves for each camera and
season with rotation periods fixed at the best estimates. The results 
are shown in Figs.~\ref{fig:lcfit-w69} and \ref{fig:lcfit-w84}. 
We then subtracted this harmonic-series fit from the original WASP lightcurves prior to our analysis of the transit.

\begin{table*} 
\caption{Frequency analysis of WASP lightcurves. 
Obs denotes WASP-South (S) or SuperWASP-North (N), 
Cam is the WASP camera number,  
N$_{\rm points}$ is the number of observations, 
$P$ is the period of corresponding to the strongest peak in the periodogram, 
Amp is the amplitude of the best-fitting sine wave in mmag and FAP is the 
false-alarm probability.} 
\label{tab:ProtTable} 
\begin{tabular}{lrrlrrrl} 
\hline 
\noalign{\smallskip}
\multicolumn{1}{l}{Year} & \multicolumn{1}{l}{Obs} & \multicolumn{1}{l}{Cam} & 
\multicolumn{1}{l}{Baseline} & 
\multicolumn{1}{l}{N$_{\rm points}$} & 
\multicolumn{1}{l}{$P$/d} & 
\multicolumn{1}{l}{Amp} & 
\multicolumn{1}{l}{FAP}\\ 
\noalign{\smallskip}
\hline
\noalign{\smallskip}
{\bf WASP-69:}\\
2008 & S & 223 & 2008 Jun 04--Oct 12 & 6732&   1.05 & 10 & $2.2 \times 10^{-2}$ \\
2008 & S & 224 & 2008 Jun 04--Oct 12 & 6003&  11.43 &  9 & $9.9 \times 10^{-4}$ \\
2008 & N & 146 & 2008 Jun 26--Oct 08 & 2928&  11.38 &  9 & $8.1 \times 10^{-4}$ \\
2009 & S & 223 & 2009 May 19--Oct 11 & 4610&  23.54 & 13 & $9.9 \times 10^{-3}$ \\
2009 & S & 224 & 2009 May 19--Oct 11 & 5111&  22.57 &  7 & $4.9 \times 10^{-4}$ \\
2009 & N & 141 & 2009 Jun 26--Oct 08 & 6556&  23.54 &  7 & $7.3 \times 10^{-5}$ \\
2009 & N & 146 & 2009 Jun 26--Oct 08 & 6677&  23.74 &  7 & $4.2 \times 10^{-6}$ \\
2010 & N & 141 & 2010 Jun 26--Oct 06 & 6846&  22.76 & 12 & $9.2 \times 10^{-10}$\\
2010 & N & 146 & 2010 Jun 26--Oct 06 & 6820&  22.76 & 11 & $7.7 \times 10^{-6}$ \\
\noalign{\smallskip}
{\bf WASP-70:} & \multicolumn{6}{l}{No evidence of modulation.}\\
\noalign{\smallskip}
{\bf WASP-84:}\\
2009 & S & 226 & 2009 Jan 14--Apr 21 & 5161  & 13.450  & 10 & $2.1 \times 10^{-9}$ \\
2010 & N & 143 & 2009 Dec 02--2010 Mar 31 & 3738  & 15.170  &  7 & $1.3 \times 10^{-4}$ \\
2010 & S & 226 & 2010 Jan 15--Apr 21 & 4548  & 15.170  &  7 & $3.7 \times 10^{-6}$ \\
2011 & N & 143 & 2010 Dec 02--2011 Mar 29 & 3701  & 14.000  & 15 & $1.8 \times 10^{-12}$ \\
2011 & S & 226 & 2011 Jan 05--Mar 01 & 3684  & 14.000  & 14 & $1.4 \times 10^{-9}$ \\
\noalign{\smallskip}
\hline 
\end{tabular} 
\end{table*} 



\section{System parameters from combined analyses}
\label{sec:syspar}
We determined the parameters of each system from a simultaneous fit to all photometric and radial-velocity data. 
The fit was performed using the current version of the 
Markov-chain Monte Carlo (MCMC) code described by \citet{2007MNRAS.380.1230C} 
and \citet{2008MNRAS.385.1576P}. 

The transit lightcurves are modelled using the formulation of 
\citet{2002ApJ...580L.171M} with the assumption that the planet is much smaller 
than the star. 
Limb-darkening was accounted for using a four-coefficient, nonlinear 
limb-darkening model, using coefficients appropriate to the passbands from the 
tabulations of \citet{2000A&A...363.1081C, 2004A&A...428.1001C}. 
The coefficients are interpolated once using the values of \logg\ and \feh\ of  
Table~\ref{tab:stars}, but are interpolated at each MCMC step using the 
latest value of \teff. The coefficient values corresponding to the best-fitting 
value of \teff\ are given in Table~\ref{tab:ld}.
The transit lightcurve is parameterised by the epoch of mid-transit 
$T_{\rm 0}$, the orbital period $P$, the planet-to-star area ratio 
(\rplanet/\rstar)$^2$, the approximate duration of the transit from initial to 
final contact $T_{\rm 14}$, and the impact parameter $b = a \cos i/R_{\rm *}$ 
(the distance, in fractional stellar radii, of the transit chord from the 
star's centre in the case of a circular orbit), where $a$ is the semimajor axis 
and $i$ is the inclination of the orbital plane with respect to the sky plane. 

\begin{table*}
\centering
\caption{Limb-darkening coefficients} 
\label{tab:ld} 
\begin{tabular}{llllllll}
\hline
\leftcell{Star}		& \leftcell{Imager}		& \leftcell{Observation bands}		& \leftcell{Claret band}			& \leftcell{$a_1$}		& \leftcell{$a_2$}		& \leftcell{$a_3$}		& \leftcell{$a_4$}	\\
\hline
WASP-69		& WASP / EulerCam	& Broad (400--700 nm) / Gunn $r$& Cousins $R$	& 0.755 & $-$0.869 & 1.581 & $-$0.633 \\
WASP-69		& TRAPPIST			& Sloan $z'$					& Sloan $z'$	& 0.824 & $-$0.922 & 1.362 & $-$0.549 \\
WASP-70A	& WASP / EulerCam	& Broad (400--700 nm) / Gunn $r$& Cousins $R$	& 0.616 & $-$0.218 & 0.743 & $-$0.397 \\
WASP-70A	& TRAPPIST			& Cousins $I$ + Sloan $z'$		& Sloan $z'$	& 0.690 & $-$0.484 & 0.828 & $-$0.402 \\
WASP-84		& WASP / RISE	& Broad (400--700 nm) / $V+R$ & Cousins $R$	& 0.695 & $-$0.591 & 1.269 & $-$0.589 \\
WASP-84		& TRAPPIST			& Sloan $z'$					& Sloan $z'$	& 0.758 & $-$0.735 & 1.162 & $-$0.517 \\
\hline
\end{tabular}
\end{table*}

The eccentric Keplerian radial-velocity orbit is parameterised by the stellar 
reflex velocity semi-amplitude $K_{\rm 1}$, the systemic velocity $\gamma$, 
an instrumental offset between the HARPS and CORALIE spectrographs
$\Delta \gamma_{\rm HARPS}$, and 
\secos\ and \sesin\,  where $e$ is orbital 
eccentricity and $\omega$ is the argument of periastron. 
We use \secos\ and \sesin\ as they impose a uniform prior on $e$, whereas 
the jump parameters we used previously, \ecos\ and \esin, impose a linear prior 
that biases $e$ toward higher values \citep{2011ApJ...726L..19A}.

The linear scale of the system depends on the orbital separation $a$ which, 
through Kepler's third law, depends on the stellar mass \mstar. 
At each step in the Markov chain, the latest values of \densstar, \teff\ and 
\feh\ are input in to the empirical mass calibration of 
\citet{2010A&A...516A..33E} (itself based on \citealt{2010A&ARv..18...67T} and 
updated by \citealt{2011MNRAS.417.2166S}) to obtain \mstar.
The shapes of the transit lightcurves and the radial-velocity curve constrain 
stellar density \densstar\ \citep{2003ApJ...585.1038S}, which combines with \mstar\ to give 
the stellar radius \rstar.
The stellar effective temperature \teff\ and metallicity \feh\ are proposal parameters constrained by Gaussian priors with 
mean values and variances derived directly from the stellar spectra 
(see Section~\ref{sec:stars}). 

As the planet-to-star area ratio is determined from the measured transit depth, 
the planet radius \rplanet\ follows from \rstar. The planet mass \mplanet\ is calculated from the 
the measured value of $K_{\rm 1}$ and the value of \mstar; the planetary density 
\densplanet\ and surface gravity $\log g_{\rm P}$ then follow. 
We  calculate the planetary equilibrium temperature \teql, assuming zero 
albedo and efficient redistribution of heat from the planet's 
presumed permanent day side to its night side. 
We also calculate the durations of transit ingress ($T_{\rm 12}$) and egress 
($T_{\rm 34}$). 

At each step in the MCMC procedure, model transit lightcurves and radial 
velocity curves are computed from the proposal parameter values, which are 
perturbed from the previous values by a small, random amount. The \chisq\ 
statistic is used to judge the goodness of fit of these models to the data and the decision as to whether to accept a step is made via the Metropolis-Hastings rule \citep{2007MNRAS.380.1230C}:
a step is accepted if \chisq\ is lower than for the previous step and a step 
with higher \chisq\ is accepted with a probability 
proportional to $\exp(-\Delta \chi^2/2)$. 
This gives the procedure some robustness against local minima and results in a thorough exploration of the parameter space around the best-fitting solution. 

For WASP-69 and WASP-84 we used WASP lightcurves detrended for rotational modulation (Section~\ref{sec:rot}). 
We excluded the WASP-69 lightcurves from camera 223 as they suffer from 
greater scatter than the lightcurves from the other cameras. 
The WASP-70A lightcurves were corrected for dilution by WASP-70B using flux ratios 
derived from in-focus EulerCam and TRAPPIST images (Section~\ref{w69-sys}). 
To give proper weighting to each photometry data set, the uncertainties were 
scaled at the start of the MCMC so as to obtain a photometric reduced-\chisq\ 
of unity. 

For WASP-69b and WASP-70Ab the improvement in the fit to the RV data resulting from the use of an eccentric orbit model is small and is consistent with the underlying 
orbit being circular. 
We thus adopt circular orbits, which \citet{2012MNRAS.422.1988A} suggest is the prudent choice for short-period,$\sim$Jupiter-mass planets in the absence of evidence to the contrary.
In a far wider orbit, closer to that of the eccentric WASP-8b 
\citep{2010A&A...517L...1Q}, WASP-84b will experience weaker tidal forces and 
there is indication that the system is young (Section~\ref{w84-sys}), so there is 
less reason to expect its orbit to be circular.
Using the $F$-test approach of \citet{1971AJ.....76..544L}, we calculate a 60 
per cent probability that the improvement in the fit could have arisen by 
chance if the underlying orbit were circular, which is similar to the probabilities for the other two systems. There is very little difference between the circular solution and the eccentric solution: the stellar density and the inferred stellar and planetary dimensions 
differ by less than half a 1-$\sigma$ error bar.
Thus we adopt a circular orbit for WASP-84b.
We find 2-$\sigma$ upper limits on $e$ of 0.10, 0.067 and 0.077 for, respectively, WASP-69b, -70Ab and -84b. 

\begin{figure*}
\begin{center}
\includegraphics[]{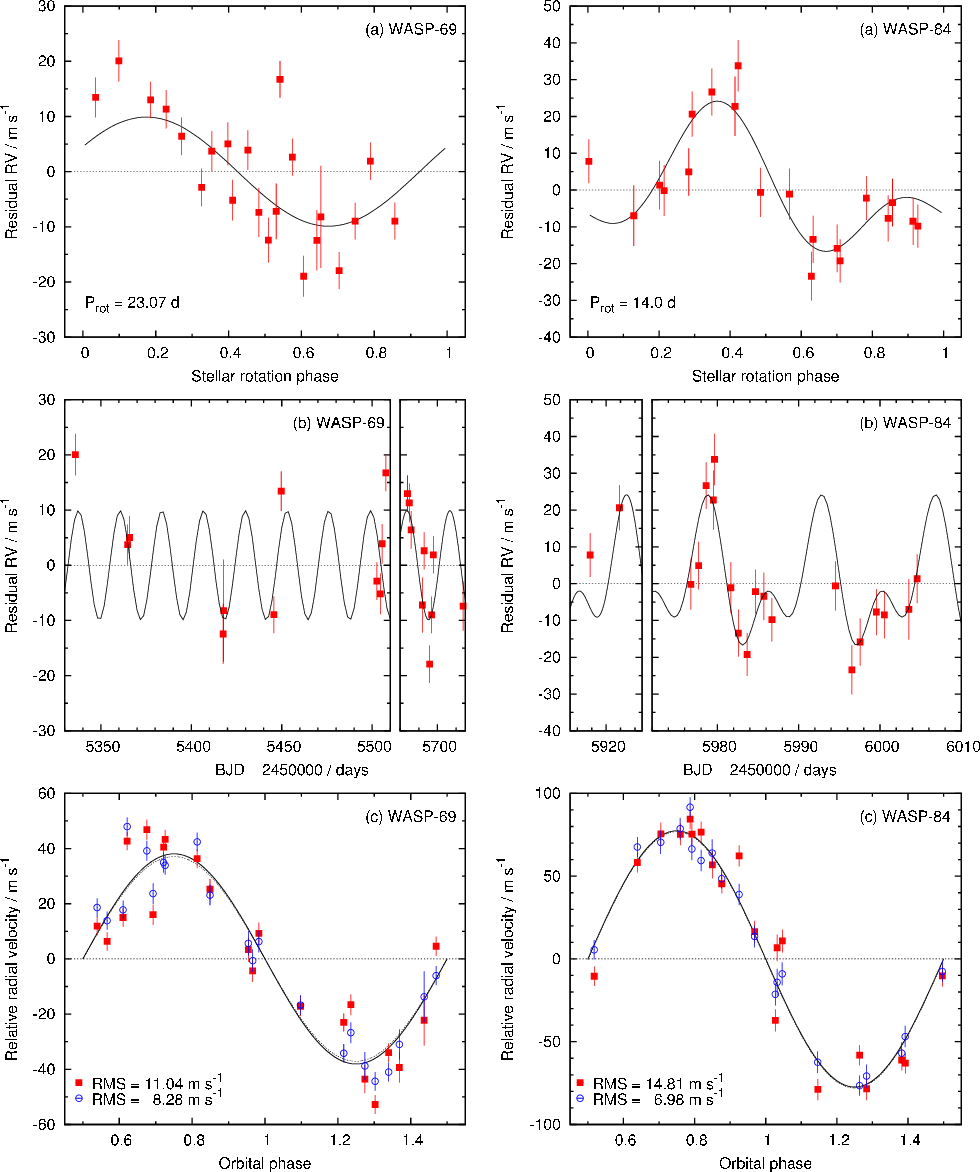}
\end{center}
\caption{The radial-velocity modulation induced by stellar activity. 
The panels on the left pertain to WASP-69 and those on the right pertain to WASP-84; 
the remainder of the caption describes the plots of both stars. 
{\bf (a)}: The residuals about the best-fitting Keplerian orbit phased on a period derived from a modulation analysis of the WASP photometry. The best-fitting harmonic function is overplotted (see Equation~\ref{sinus} and Table~\ref{tab:act-rv}). 
{\bf (b)}: The residuals about the best-fitting Keplerian orbit as a function of time; the best-fitting harmonic function is overplotted.  The abscissa scales on the two adjacent panels are equal. 
Specific to WASP-69, two points are not plotted; one was taken a season earlier and the other a season later than the data shown.
With reference to the top-left panel (panel (a) for WASP-69), the `early' point is at coordinate (0.61,$-$19) and the `late' point is at coordinate (0.51,$-$12). 
{\bf (c)}: The RVs, both detrended and non-detrended, folded on the transit ephemeris. By first detrending the RVs with the harmonic function a significantly lower scatter is obtained (the blue circles about the dashed line) as compared to the non-detrended RVs (the red squares about the solid line). 
\label{fig:rv-act} }
\end{figure*}

From initial MCMC runs we noted excess scatter in the RV residuals of both WASP-69 and WASP-84, with the residuals varying sinusoidally when phased on the stellar rotation periods, as derived from the WASP photometry (Figure~\ref{fig:rv-act}). 
The RV residuals of WASP-69 are adequately fit by a sinusoid and those of WASP-84 benefit from the addition of the first harmonic:

\begin{equation}
  RV_{\rm activity} = a_1 \sin 2\pi\phi + 
                 b_1 \cos 2\pi\phi +
                 a_2 \sin 4\pi\phi.
\label{sinus}
\end{equation} 

\noindent We phased the RV residuals of WASP-69, which were obtained over two seasons, with the average rotation period (23.07\,d) derived from all available WASP photometry. For WASP-84 we chose to phase the RV residuals, which were obtained in a single season, with the period derived from the photometry obtained in the preceding season (2011; $P_{\rm rot}$ = 14.0\,d), as we did not perform simultaneous monitoring. 
Pre-whitening the RVs using Equation~\ref{sinus} reduced the RMS of the residual scatter about the best-fitting Keplerian orbit from 11.04\,\ms\ to 8.28\,\ms\ in the case of WASP-69 and from 14.81\,\ms\ to 6.98\,\ms\ in the case of WASP-84 (Figure~\ref{fig:rv-act}). 
The coefficients of the best-fitting activity models are given in Table~\ref{tab:act-rv}.

\begin{figure*}
\begin{center}
\includegraphics[]{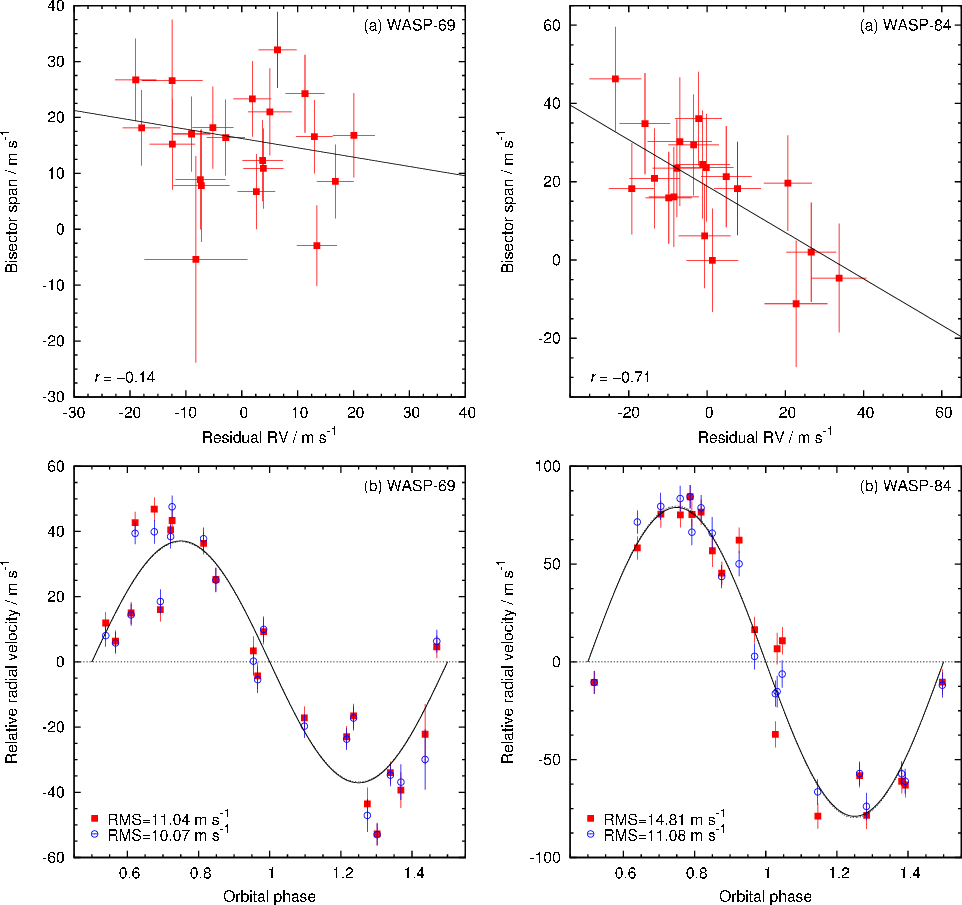}
\end{center}
\caption{The anticorrelated variations in bisector span and radial velocity induced by stellar activity. 
From initial MCMCs for both WASP-69 and WASP-84, we noted large scatter about the best-fitting Keplerian orbits, which we suggest to be the product of stellar activity. 
The panels on the left pertain to WASP-69 and those on the right pertain to WASP-84; 
the remainder of the caption describes the plots of both stars. 
{\bf (a)}: Similar to \citet{2007A&A...467..721M}, we found the residuals to be anticorrelated with bisector span. A linear fit to the data is overplotted and the correlation coefficient, $r$ is given. 
{\bf (b)}: By first detrending the RVs with this bisector-residual relation, a smaller scatter about the best-fitting Keplerian orbit is obtained (the blue circles about the dashed line) as compared to the non-detrended RVs (the red squares about the solid line). 
The result is inferior to that obtained when detrending the RVs with low-order harmonic series  (Figure~\ref{fig:rv-act}).
\label{fig:resid-bis} }
\end{figure*}

A common alternative approach is to pre-whiten RVs using a relation derived from a linear fit between the RV residuals and the bisector spans, which are anticorrelated when stellar activity is the source of the signal \citep{2007A&A...467..721M}. 
The WASP-69 RV residuals are essentially uncorrelated with the bisector spans ($r=-0.14$) resulting in a small reduction in the RMS of the residual RVs, from 11.04\,\ms\ to 10.08\,\ms (see Figure~\ref{fig:resid-bis}). 
This is probably due to the star's low \vsini\ (2.2\,\kms) as  
RV varies linearly with \vsini, but bisector span goes as (\vsini)$^{3.3}$
\citep{1997ApJ...485..319S,2003A&A...406..373S}. 
The anticorrelation ($r = -0.71$) between the RV residuals and the bisector spans of WASP-84 (\vsini\ = 4.1\,\kms) is strong and the RMS of the residual RVs is reduced from 14.81\,\ms\ to 11.08\,\ms (see Figure~\ref{fig:resid-bis}). 
The reduction in the scatter of the residual RVs obtained using this approach was less for both WASP-69 and WASP-84 than that obtained using the harmonic-fitting approach; we thus elected to use the RVs pre-whitened by subtracting harmonic functions in the MCMC analyses.

For both WASP-69 and WASP-84 the best-fitting solution is essentially unchanged by the pre-whitening. 
One parameter that we could expect to be affected, via a modified $K_{\rm 1}$, is the planetary mass. 
For WASP-69b we obtain \mplanet~$= 0.260 \pm 0.017$\,\mjup\ when using pre-whitened RVs and 
\mplanet $= 0.253 \pm 0.022$\,\mjup\ otherwise.
For WASP-84b we obtain \mplanet~$= 0.694 \pm 0.028$\,\mjup\ when using pre-whitened RVs and 
\mplanet~$= 0.691^{+0.061}_{-0.032}$\,\mjup\ otherwise.

To obtain a spectroscopic reduced $\chi^2$ of unity we added a `jitter' term in quadrature 
to the formal RV errors. 
Pre-whitening reduced the jitter required from 10.1\,\ms\ to 6.7\,\ms\ for WASP-69 and from 13.5\,\ms\ to zero for WASP-84; the jitter for WASP-70A was 3.9\,\ms.
We excluded two in-transit WASP-70A RVs from the analysis as we did not model 
the Rossiter-McLaughlin effect \citep[e.g.][]{2012MNRAS.423.1503B}. 

We explored the possible impact of spots on our determination of the system parameters of WASP-69 and WASP-84. 
We assumed that the stars' visible faces were made permanently dimmer, by spots located outside of the transit chords, at the level of the maximum amplitudes of the rotational-modulation signals found from the WASP photometry (Table~\ref{tab:ProtTable}). 
Having corrected the transit lightcurves for this, we performed MCMC analyses again and found the stellar and planetary dimensions (density, radius, mass) to have changed by less than 0.2\,$\sigma$. 

\begin{table}
\centering
\caption{Coefficients of the activity-induced RV variations model}
\label{tab:act-rv}
\begin{tabular}{lrrr}
\hline
Star    & \leftcell{$a_1$}	& \leftcell{$b_1$}	& \leftcell{$a_2$}	\\
        & \leftcell{(\ms)}	& \leftcell{(\ms)}	& \leftcell{(\ms)}	\\
\hline
WASP-69 & $8.8 \pm 2.6$ & $4.5 \pm 2.7$ & 0		\\
WASP-84 & $12.1 \pm 2.3$& $-6.5 \pm 2.8$& $-10.9 \pm 2.5$ \\
\hline
\end{tabular}
\end{table}

The system parameters from the combined analyses are given in 
Table~\ref{tab:mcmc} and the best fits to the radial velocities and the photometry 
are plotted in 
Figures~\ref{fig:w69-rv-phot}, \ref{fig:w70-rv-phot} and \ref{fig:w84-rv-phot}.

\begin{table*} 
\centering
\caption{System parameters} 
\label{tab:mcmc}
\begin{tabular}{lcccc}
\hline
Parameter & Symbol (Unit) & WASP-69b & WASP-70Ab & WASP-84b\\ 
\hline 
\\
Orbital period & $P$ (d) & $3.8681382 \pm 0.0000017$ & $3.7130203 \pm 0.0000047$ & $8.5234865 \pm 0.0000070$ \\
Epoch of mid-transit & $T_{\rm c}$ (BJD, UTC) & $2455748.83344 \pm 0.00018$ & $2455736.50348 \pm 0.00027$ & $2456286.10583 \pm 0.00009$ \\
Transit duration & $T_{\rm 14}$ (d) & $0.0929 \pm 0.0012$ & 0.1389$^{+ 0.0014}_{- 0.0019}$ & $0.11452 \pm 0.00046$ \\
Transit ingress/egress duration & $T_{\rm 12}=T_{\rm 34}$ (d) & $0.0192 \pm 0.0014$ & 0.0150$^{+ 0.0016}_{- 0.0021}$ & $0.02064 \pm 0.00060$ \\
Planet-to-star area ratio & $\Delta F=R_{\rm P}^{2}$/R$_{*}^{2}$ & $0.01786 \pm 0.00042$ & 0.00970$^{+ 0.00022}_{- 0.00026}$& $0.01678 \pm 0.00015$ \\
Impact parameter & $b$ & $0.686 \pm 0.023$ & 0.431$^{+ 0.076}_{- 0.169}$ & $0.632 \pm 0.012$ \\
Orbital inclination & $i$ (\deg) \medskip & $86.71 \pm 0.20$ & 87.12$^{+ 1.24}_{- 0.65}$ & $88.368 \pm 0.050$ \\
Stellar reflex velocity semi-amplitude & $K_{\rm 1}$ (\ms) & $38.1 \pm 2.4$ & $72.3 \pm 2.0$ & $77.4 \pm 2.0$ \\
Systemic velocity & $\gamma$ (\ms)  & $-9\,628.26 \pm 0.23$ & $-65\,389.74 \pm 0.32$ & $-11\,578.14 \pm 0.33$ \\
Offset between CORALIE and HARPS & $\Delta \gamma_{\rm HARPS}$ (\ms) \medskip & --- & $15.44 \pm 0.11$ & --- \\
Eccentricity & $e$ & 0 (adopted) ($<$0.10 at 2\,$\sigma$) & 0 (adopted) ($<$0.067 at 2\,$\sigma$) & 0 (adopted) ($<$0.077 at 2\,$\sigma$)\\
Stellar mass & $M_{\rm *}$ ($M_{\rm \odot}$) & $0.826 \pm 0.029$ & $1.106 \pm 0.042$ & $0.842 \pm 0.037$ \\
Stellar radius & $R_{\rm *}$ ($R_{\rm \odot}$) & $0.813 \pm 0.028$ & 1.215$^{+ 0.064}_{- 0.089}$ & $0.748 \pm 0.015$ \\
Stellar surface gravity & $\log g_{*}$ (cgs) & $4.535 \pm 0.023$ & 4.314$^{+ 0.052}_{- 0.036}$& $4.616 \pm 0.011$ \\
Stellar density & $\rho_{\rm *}$ ($\rho_{\rm \odot}$) & $1.54 \pm 0.13$ & 0.619$^{+ 0.136}_{- 0.077}$ & $2.015 \pm 0.070$ \\
Stellar effective temperature & $T_{\rm eff}$ (K) & $4715 \pm 50$ & $5763 \pm 79$ & $5314 \pm 88$\\
Stellar metallicity & {[Fe/H]} \medskip & $0.144 \pm 0.077$ & $-0.006 \pm 0.063$ & $0.00 \pm 0.10$\\
Planetary mass & $M_{\rm P}$ ($M_{\rm Jup}$) & $0.260 \pm 0.017$ & $0.590 \pm 0.022$ & $0.694 \pm 0.028$ \\
Planetary radius & $R_{\rm P}$ ($R_{\rm Jup}$) & $1.057 \pm 0.047$ & 1.164$^{+ 0.073}_{- 0.102}$ & $0.942 \pm 0.022$ \\
Planetary surface gravity & $\log g_{\rm P}$ (cgs) & $2.726 \pm 0.046$ & 3.000$^{+ 0.066}_{- 0.050}$& $3.253 \pm 0.018$ \\
Planetary density & $\rho_{\rm P}$ ($\rho_{\rm J}$) & $0.219 \pm 0.031$ & 0.375$^{+ 0.104}_{- 0.060}$& $0.830 \pm 0.048$ \\
Orbital major semi-axis & $a$ (AU)  & $0.04525 \pm 0.00053$ & $0.04853 \pm 0.00062$ & $0.0771 \pm 0.0012$ \\
Planetary equilibrium temperature & \teql\ (K) & $963 \pm 18$ & $1387 \pm 40$ & $797 \pm 16$ \\
\\ 
\hline 
\end{tabular} 
\end{table*}

\section{The WASP-69 system}
\label{w69-sys}
WASP-69b is a bloated Saturn-mass planet (0.26\,\mjup, 1.06\,\rjup) in a 3.868-d orbit around 
an active mid-K dwarf. 
The system is expected to be a favourable target for transmission spectroscopy owing to the large predicted scale height of the planet's atmosphere and the apparent brightness and the small size of the star.
The parameters derived from the spectral analysis and the MCMC analysis are given, 
respectively, in Tables~\ref{tab:stars} and \ref{tab:mcmc} and the corresponding transit and Keplerian orbit models are superimposed, respectively, on the radial velocities and the photometry in Figure~\ref{fig:w69-rv-phot}. 

We estimated the stellar rotation period $P_{\rm rot}$ from activity-rotation induced modulation of the WASP lightcurves to be $23.07\pm0.16$\,d. 
Together with our estimate of the  stellar radius (Table~\ref{tab:mcmc}), this 
implies a rotation speed of $v = 1.78 \pm 0.06$\,\kms, which can be compared with 
the spectroscopic estimate of the projected rotation speed of 
\vsini\ = $2.2 \pm 0.4$\,\kms. 

The rotation rate ($P \leq 18.7 \pm 3.5$~d) implied by the {\vsini} gives a
gyrochronological age of $0.73 \pm 0.28$~Gyr using the
\citet{2007ApJ...669.1167B} relation. The lightcurve-modulation period implies a slightly older gyrochronological age of $1.10\pm0.15$~Gyr.

There is no significant detection of lithium in the spectra of WASP-69, with an 
equivalent-width upper limit of 12\,m\AA, corresponding to an abundance upper 
limit of $\log A$(Li) $<$ 0.05 $\pm$ 0.07. This implies an age of at least 
0.5~Gyr \citep{2005A&A...442..615S}.

\begin{figure}
\includegraphics[width=\columnwidth]{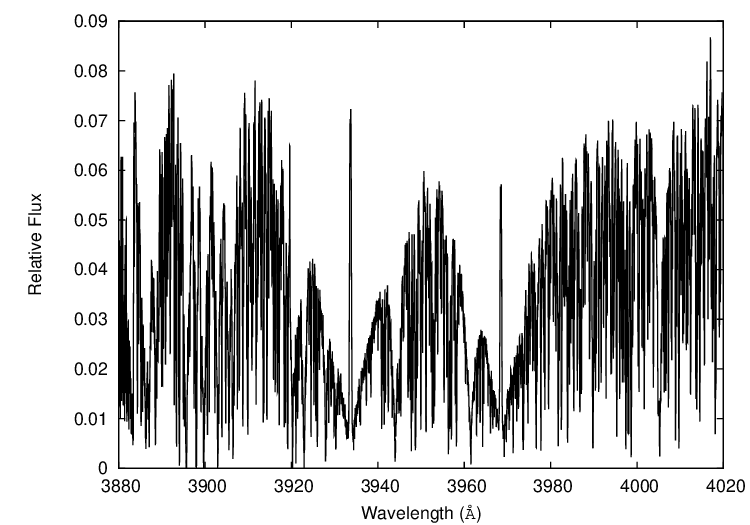}
\caption{CORALIE spectrum of WASP-69 showing strong emission in the Ca~{\sc ii} H+K line cores.}
\label{H+K}
\end{figure}

There are strong emission peaks evident in the Ca~{\sc ii} H+K lines
(Figure~\ref{H+K}), with an estimated activity index of \rhk~$\sim
-4.54$. This gives an approximate age of $\sim$0.8~Gyr according to
\citet{2008ApJ...687.1264M}, which is consistent with the age implied from the
rotation rate and the absence of lithium.

Interestingly, the observed rotation period and the 2MASS $J-K$ value of 0.57 
suggest an age closer to 3~Gyr if we apply $P \simeq \sqrt{t}$ using the Coma Berenices
cluster colour-rotation distribution as a benchmark \citep{2009MNRAS.400..451C}.
This is consistent with the lack of lithium, but slightly at odds with the 
\rhk\ and the \citet{2007ApJ...669.1167B} gyrochronological ages.

As a low-density planet in a short orbit around a relatively-young, active star, WASP-69b could be undergoing significant mass loss due to X-ray-driven or extreme-ultraviolet-driven evaporation (e.g. \citealt{2007A&A...461.1185L,2012MNRAS.422.2024J}). 
At earlier times the stellar X-ray luminosity, and hence the planetary mass loss rate, would have been higher.  
ROSAT recorded an X-ray source 1RXS~J210010.1$-$050527 with a count rate of $2.4\pm0.9$\,s$^{-1}$ over 0.1--2.4\,keV and at an angular distance of 60$\pm$27\arcsec\ from WASP-69. 
There is no optical source spatially coincident with the ROSAT source position. The brightest object in the NOMAD catalog located within 60\arcsec\ of the ROSAT source is 30\arcsec\ away and has R=16.6, whereas WASP-69 is at 60\arcsec\ and has R=9.2. 
Assuming a uniform sky distribution for the 18\,811 bright sources (count rate $>0.05$\,s$^{-1}$) of the ROSAT all-sky catalogue \citep{1999A&A...349..389V}, the probability that an unrelated source is within 120\arcsec\ of WASP-69 is 0.05 per cent. 

Thus, assuming the ROSAT source to be WASP-69, we follow \citet{2012MNRAS.422.2024J} to estimate a current planetary mass loss rate of 10$^{12}$\,g\,s$^{-1}$, having assumed an evaporation efficiency factor of 0.25 and having converted the count rate to an X-ray luminosity ($\log (L_{\rm X}/L_{\rm bol}) = -4.43$; \citealt{1995ApJ...450..401F}).
This can be compared with the mass-loss rates of HD\,209458b and HD\,189733b, estimated at $\sim$10$^{10\mbox{--}11}$\,g\,s$^{-1}$ from observations of Lyman $\alpha$ absorption by escaping hydrogen atoms \citep{2003Natur.422..143V,2010A&A...514A..72L}. 
\citet{2012MNRAS.422.2024J} estimate that a star of similar spectral type as WASP-69 has a saturated X-ray luminosity ratio of $\log (L_{\rm X}/L_{\rm bol})_{\rm sat} \sim -3.35$ for the first $\sim$200\,Myr of its life. Assuming WASP-69b to have been in-situ during this period suggests the planet would have lost mass at a rate of $\sim$10$^{13}$\,g\,s$^{-1}$ and, assuming a constant planetary density, the planet to have undergone a fractional mass loss of $\sim$0.2. 

\section{The WASP-70 system}
\label{w70-sys}
WASP-70Ab is a 0.59\mjup\ planet in a 3.713-d orbit around the primary of a 
spatially-resolved G4+K3 binary. 
The parameters derived from the spectral analysis and the MCMC analysis are given, 
respectively, in Tables~\ref{tab:stars} and \ref{tab:mcmc} and the corresponding transit and Keplerian orbit models are superimposed, respectively, on the radial velocities and the photometry in Figure~\ref{fig:w70-rv-phot}. 

\begin{figure}
\includegraphics[height=\columnwidth]{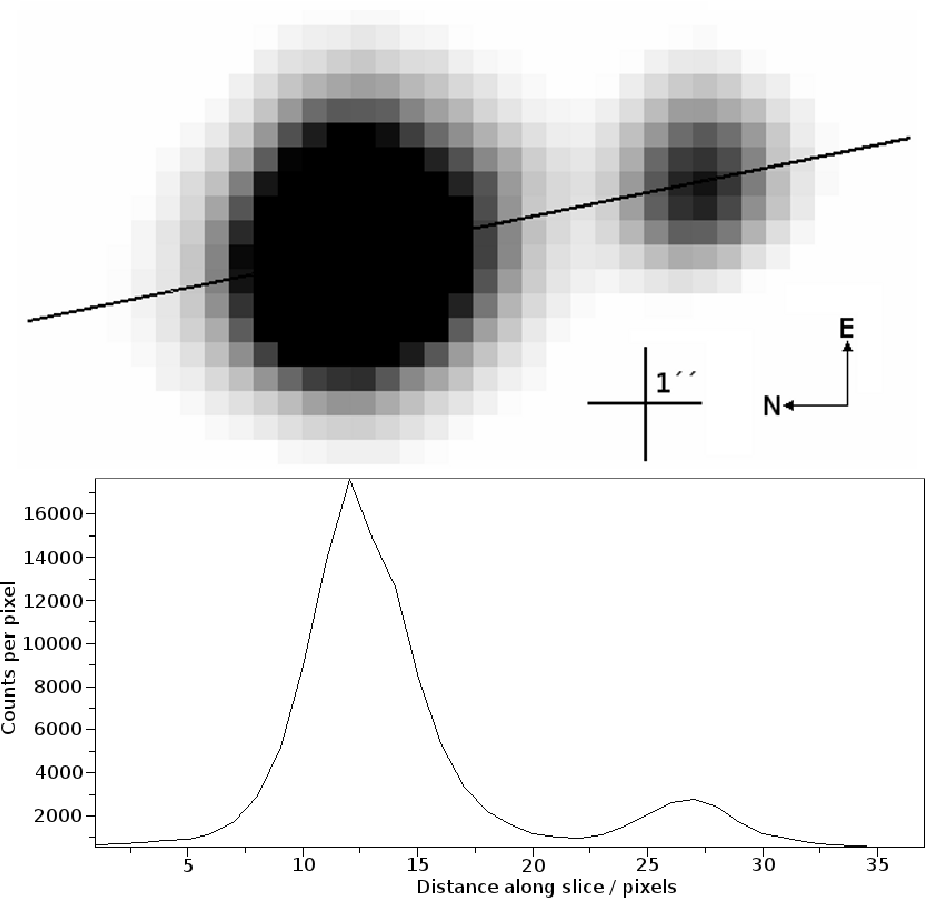}
\caption{{\it Upper portion}: EulerCam Gunn-$r$ image of WASP-70A+B; 
the B component is at a distance of 3\farcs3 and position angle of 167\degr. 
{\it Lower portion}: The counts per pixel along the slice marked on the image.
\label{fig:w70ab}}
\end{figure}

2MASS images reveal that 
the two stars are separated by $\sim$3\arcsec\ with a position angle of 167\degr.
We obtained in-focus TRAPPIST $I+z'$ and EulerCam Gunn $r$ images in 2011 and 2012, respectively (Figure~\ref{fig:w70ab}). 
These show the two stars separated by 3\farcs3 at a position angle of 167\degr, 
which is consistent with no significant relative motion since the 2MASS images 
were taken in 1998.
The projected separation of 3\farcs3 and inferred distance of 245~pc,
give a physical separation of the two stars of at least 800~AU.
Further evidence of the stars' association comes from the mean radial velocities 
measured from  CORALIE spectra: $-$65.4\,\kms\ and  
$-$64.6\,\kms\ for WASP-70A and WASP-70B, respectively. 

\begin{table}
\caption{Estimated magnitudes and flux ratios of WASP-70A+B 
\label{tab:mags}}
\centering
\begin{tabular}{ccccc} \hline
Band & WASP-70A & WASP-70B &  $\Delta_{\rm A-B}$ & $f_{\rm B}/f_{\rm A}$\\ 
\hline
$r$    & 10.8  & 13.1  & $-2.3$ & 0.12\\
$I+z'$ & \ldots & \ldots & $-1.6$ & 0.22\\ 
\hline
\end{tabular}
\end{table}

\begin{figure}
\includegraphics[width=\columnwidth]{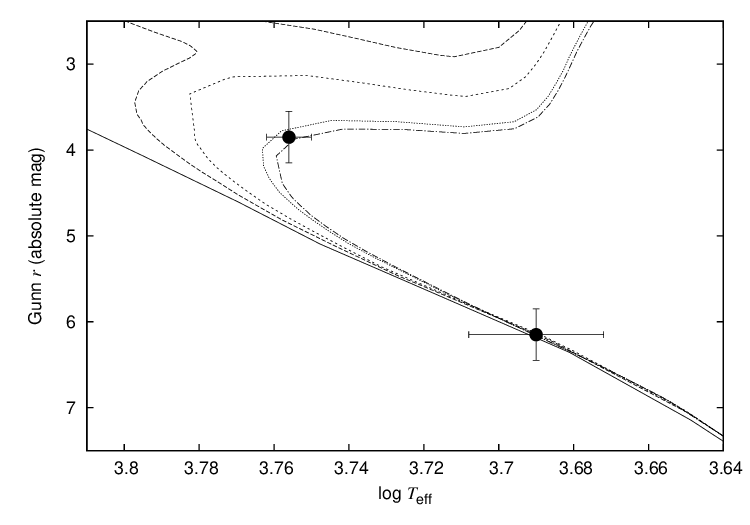}
\caption{H-R diagram for WASP-70A+B.
Various isochrones from \citet{2008A&A...482..883M} are plotted, with ages of 
0, 3, 5, 9 and 10 Gyr.
\label{w70-hr}}
\end{figure}

We determined the flux ratios of the stellar pair from aperture photometry of the images. For the EulerCam images, we used the USNO-B1.0 magnitudes of comparison stars to determine 
approximate $r$-band magnitudes (Table~\ref{tab:mags}). 
By combining these with the \teff\ from the spectral analysis we
constructed a Hertzsprung-Russell diagram for the WASP-70 double system (Figure~\ref{w70-hr}). 
A distance modulus of $6.95 \pm 0.15$ ($\equiv 245 \pm 20$~pc) was required to bring
the companion star on to the main sequence. WASP-70A appears to have evolved off
the ZAMS with an age of around 9--10\,Gyr.

The rotation rate for WASP-70A ($P \leq 34.1 \pm 7.8$~d) implied by the {\vsini}
gives a gyrochronological age of $\leq 7 \pm 3$~Gyr using the
\citet{2007ApJ...669.1167B} relation.
There is no significant detection of lithium in the spectra, with 
equivalent-width upper limits of 11m{\AA} and 20m{\AA}, corresponding to 
abundance upper limits of $\log A$(Li) $<$ 1.20 $\pm$ 0.07 and 
$<$ 0.71 $\pm$ 0.25 for WASP-70A and B, respectively. These imply an age of at 
least $\sim$5 Gyr \citep{2005A&A...442..615S} for WASP-70A and over 600\,Myr for 
WASP-70B.
We found no evidence of modulation in the WASP-70A+B lightcurves or of emission 
peaks in the Ca~{\sc ii} H+K lines in either star.

\section{The WASP-84 system}
\label{w84-sys}
WASP-84b is a 0.69-\mjup\ planet in an 8.523-d orbit around an early-K dwarf. 
After HAT-P-15b and HAT-P-17b, which have orbital periods of 10.86\,d and 10.34\,d, respectively, WASP-84b has the longest orbital period of the planets discovered from the ground by the transit technique \citep{2010ApJ...724..866K, 2012ApJ...749..134H}. 
The parameters derived from the spectral analysis and the MCMC analysis are given, 
respectively, in Tables~\ref{tab:stars} and \ref{tab:mcmc} and the corresponding transit and Keplerian orbit models are superimposed, respectively, on the radial velocities and the photometry in Figure~\ref{fig:w84-rv-phot}. 

We estimated the stellar rotation period from the modulation of the WASP 
lightcurves to be $P_{\rm rot}$ = $14.36\pm0.35$\,d. 
Together with our estimates of the  stellar radius (Table~\ref{tab:mcmc}), this 
implies a rotation speed of $v = 2.64 \pm 0.08$\,\kms. This is inconsistent with 
the spectroscopic estimate of the projected rotation speed of 
\vsini = $4.1 \pm 0.3$\,\kms. 
This disagreement may be due to an overestimation of \vsini\ and an underestimation of \rstar. 
We took macroturbulence values from the tabulation of \cite{2010MNRAS.405.1907B}. 
If we instead adopted a higher macroturbulence value from \citet{2008oasp.book.....G} 
then we would obtain a lower \vsini\ of $\sim$3.5\,\kms. 
Similarly, systematics in the lightcurves or a limitation of the empirical mass 
calibration could have resulted in an underestimation of the stellar radius.

Using the \citet{2007ApJ...669.1167B} relation, $P_{\rm rot}$ gives a gyrochronological age of $0.79 \pm 0.12$~Gyr, whereas the rotation rate implied by the \vsini\ ($P \leq 9.2 \pm 0.7$\,d) gives $\leq$$0.34 \pm 0.07$~Gyr. 

There is no significant detection of lithium in the spectra, with an equivalent
width upper limit of 8m\AA, corresponding to an abundance upper limit of $\log
A$(Li) $<$ 0.12 $\pm$ 0.11. This implies an age of at least $\sim$0.5~Gyr \citep{2005A&A...442..615S}.

There are emission peaks evident in the Ca~{\sc ii} H+K lines, with an estimated activity index of \rhk=$\sim -4.43$. This gives an approximate age of 0.4\,Gyr according to \citet{2008ApJ...687.1264M}.

Using the Coma Berenices cluster colour-rotation relation of 
\citet{2009MNRAS.400..451C} we find an age of 1.4~Gyr. As was the case with WASP-69, this is at odds with the \rhk\ and the \citet{2007ApJ...669.1167B} gyrochronological ages.

\section{Discussion and summary}
\label{summary}
We reported the discovery of the transiting exoplanets WASP-69b, WASP-70Ab and WASP-84b, each of which orbits a bright star ($V\sim10)$. 
We derived the system parameters from a joint analysis of the WASP survey photometry, the CORALIE and HARPS radial velocities and the high-precision photometry from TRAPPIST, EulerCam and RISE. 

WASP-69b is a bloated Saturn-type planet (0.26\,\mjup, 1.06\,\rjup) in a 3.868-d period around a mid-K dwarf. 
We find WASP-69 to be active from emission peaks in the Ca~{\sc ii} H+K lines (\rhk~$\sim -4.54$) and from modulation of the star's lightcurves ($P_{\rm rot} = 23.07 \pm 0.16$ d; amplitude = 7--13 mmag). 
Both the gyrochronological calibration of \citet{2007ApJ...669.1167B} and the age-activity relation of \citet{2008ApJ...687.1264M} suggest an age of around 1\,Gyr, whereas the Coma Berenices colour-rotation calibration of \citep{2009MNRAS.400..451C} suggests an age closer to 3\,Gyr. 

WASP-70Ab is a Jupiter-type planet (0.59\,\mjup, 1.16\,\rjup) in a 3.713-d orbit around the primary of a spatially-resolved G4+K3 binary separated by 3\farcs3 ($\geq$800\,AU). 
An absence of emission in the Ca~{\sc ii} H+K lines and an absence of lightcurve modulation indicates WASP-70A is relatively inactive.  
We used the binary nature of the system to construct an H-R diagram, from which we estimate its age to be 9--10\,Gyr.

WASP-84b is a Jupiter-type planet (0.69\,\mjup, 0.94\,\rjup) in an 8.523-d orbit around an active early-K dwarf. The planet has the third-longest period of the transiting planets discovered from the ground. 
We find WASP-84 to be active from emission peaks in the Ca~{\sc ii} H+K lines (\rhk~$\sim -4.43$) and from modulation of the star's lightcurves ($P = 14.36 \pm 0.35$ d; amplitude = 7--15 mmag). 
The gyrochronological calibration of \citet{2007ApJ...669.1167B} suggests an age of $\sim$0.8\,Gyr for WASP-84, whereas the age-activity relation of \citet{2008ApJ...687.1264M} suggests an age of $\sim$0.4\,Gyr.

We used the empirical relations of \citet{2012A&A...540A..99E}, derived from fits to the properties of a well-characterised sample of transiting planets, to predict the radii of the three presented planets. 
The difference between the observed radii (Table~\ref{tab:mcmc}) and the predicted radii of both WASP-70Ab ($\Delta R_{\rm P,obs-pred}$ = $-$0.067\,\rjup) and WASP-84b ( ($\Delta R_{\rm P,obs-pred}$ = 0.065\,\rjup)) are small. 
For comparison, the average difference between the predicted and observed radii for the sample of \citet{2012A&A...540A..99E} is 0.11\,\rjup. 
The predicted radius of WASP-69b (0.791\,\rjup) is smaller than the observed radius (Table~\ref{tab:mcmc}) by 0.266\,\rjup, showing the planet to be bloated. 

We observed excess scatter in the radial-velocity (RV) residuals about the best-fitting Keplerian orbits for the two active stars WASP-69 and WASP-84. 
Phasing the RV residuals on the photometrically-determined stellar rotation periods showed the excess scatter to be induced by a combination of stellar activity and rotation. 
We fit the residuals with low-order harmonic series and subtracted the best fits from the RVs prior to deriving the systems' parameters. The systems' solutions were essentially unchanged by this, with much less than a 1-$\sigma$ change to the planet mass in each case. 
We found this method of pre-whitening using a harmonic fit to result in a greater reduction in the residual RV scatter ($\Delta$RMS = $-$2.76\,\ms\ for WASP-69 and $\Delta$RMS = $-$7.83\,\ms\ for WASP-84) than the more traditional method of pre-whitening with a fit to the RV residuals and the bisector spans ($\Delta$RMS = $-$0.96\,\ms\ for WASP-69 and $\Delta$RMS = $-$3.73\,\ms\ for WASP-84). 
The poor performance of the bisector-RV fit method for WASP-69 probably stems from the slow rotation of the star, which has been shown to limit the method's efficacy \citep{1997ApJ...485..319S, 2003A&A...406..373S}.

WASP-69 and WASP-84 are two of the most active stars known to host exoplanets. 
Of the 303 systems with measured \rhk\ values only 17 are more active than WASP-69 and only 6 are more active than WASP-84\footnote{The data were retrieved from \url{http://exoplanets.org} on 2013 May 5. See \citet{2011PASP..123..412W}.}. Conversely, only 14 exoplanet host stars have measured \rhk\ indices indicating lower activity levels than WASP-70A, though values for inactive stars are probably underreported. 

\section*{Acknowledgements}
WASP-South is hosted by the South African Astronomical Observatory and SuperWASP-North is hosted by the Isaac Newton Group on La Palma. We are grateful for their ongoing support and assistance. 
Funding for WASP comes from consortium universities and from the UK's Science and Technology Facilities Council. TRAPPIST is funded by the Belgian Fund for Scientific Research (Fond National de la Recherche Scientifique, FNRS) under the grant FRFC 2.5.594.09.F, with the participation of the Swiss National Science Fundation (SNF). 
The Liverpool Telescope is operated on the island of La Palma by Liverpool John Moores University in the Spanish Observatorio del Roque de los Muchachos of the Instituto de Astrofisica de Canarias with financial support from the UK Science and Technology Facilities Council. 
M. Gillon and E. Jehin are FNRS Research Associates.
A. H.M.J. Triaud is a Swiss National Science Foundation fellow under grant number PBGEP2-145594.
L. Delrez is a FNRS/FRIA Doctoral Fellow. 
We are grateful to R.~D.~Jeffries for discussion of the ROSAT catalog. 
This research has made use of the Exoplanet Orbit Database and the Exoplanet Data Explorer at \url{http://exoplanets.org} and the VizieR catalogue access tool, CDS, Strasbourg, France. The original description of the VizieR service was published in A\&AS 143, 23.


\label{lastpage}

\end{document}